\documentclass[conference]{IEEEtran}
\usepackage{subfigure}
\usepackage{multirow} 
\usepackage{booktabs}
\usepackage{tabularx}
\usepackage[ruled,linesnumbered]{algorithm2e}
\usepackage{pifont}
\usepackage{adjustbox}
\usepackage{tcolorbox}
\usepackage{seqsplit}
\usepackage{enumitem}
\usepackage{amsmath} 
\usepackage{amssymb}
\usepackage{marvosym} 
\usepackage{url} 
\usepackage{xurl}
\usepackage[hidelinks]{hyperref}

\ifCLASSINFOpdf
\else
\fi
\begin{document}


\title{From Instructions to Exploits: Applying Program Analysis to Malicious Agent Skills Across Natural Language and Code}


\author{%
\begin{minipage}{\textwidth}
\centering

{
Shiwu Zhao$^{1,*}$,
Baizhu Wang$^{2,*}$,
Guoqiang Zheng$^{1}$,
Tingyu Gao$^{2}$,
Yuheng Xie$^{1}$,
Rujie Chen$^{2}$,
Xingchen Chen$^{1,\dagger}$,
Xuezhang Liu$^{2}$,
Qixu Liu$^{1}$
\par}

{
$^{1}$Institute of Information Engineering, CAS, Beijing, China\\
$^{2}$MYbank, Ant Group, Hangzhou, China\\

\texttt{\{zhaoshiwu,zhengguoqiang,xieyuheng,chenxingchen,liuqixu\}@iie.ac.cn}\\
\texttt{\{wangbaizhu.wbz,gaotingyu.gty,chenrujie.crj,liuxuezhang.lxz\}@mybank.cn}\\

$^{*}$These authors contributed equally to this work.\\
$^{\dagger}$Corresponding author
}
\end{minipage}
}

\maketitle

\begin{abstract}

LLM-powered agents increasingly rely on third-party skills to gain tool-use capabilities, turning agent skill marketplaces into a new supply-chain attack surface. 
Unlike traditional software packages, these skills are hybrid artifacts where natural-language instructions guide execution logic while bundled code performs specific tasks, and malicious behavior can emerge from interactions between these two artifact types.
Existing skill-security detection methods either rely heavily on rule matching or depend excessively on LLM-based judgments. 
Meanwhile, they provide limited support for capturing malicious behavior that spans natural-language instructions and executable code.

To address these challenges, we present the \textit{Supply-Chain Scanner for Agent Skills} (\textit{SCS}), a static analysis framework that brings natural-language instructions and bundled code into a unified program-analysis pipeline.
\textit{SCS} compiles natural-language instructions into a Python-style intermediate representation (IR), based on a limited set of APIs that we have defined. 
This intermediate representation, together with the bundled code, forms a unified cross-domain Program Dependence Graph (xPDG).
On top of xPDG, we introduce TSC (Trust--Sensitivity--Capability), a three-dimensional security property model that propagates data provenance, sensitivity, and capability information over variables.
We define deterministic TSC propagation and rule-based downstream analysis over the unified xPDG, enabling structured cross-domain analysis while reducing reliance on direct LLM-based security judgments.

For evaluation, we construct a manually verified benchmark containing 117 malicious and 117 benign skills.
Across five LLMs, SCS achieves the highest average accuracy (85.6\%), recall (95.2\%), and F1-score (86.9\%) among the evaluated methods, with a precision of 80.2\% and average token consumption of 13.7M.
Its NL-to-IR compiler consistently compiles over 92\% of human-annotated instruction steps across different LLMs.
Finally, we collect 66,157 real-world ClawHub skills and, using SCS together with auxiliary dynamic validation and manual auditing, confirm 627 malicious skills.


\end{abstract}


%
\IEEEpeerreviewmaketitle

\section{Introduction}
\label{sec:intro}

With the rapid development of the large language model (LLM) agent ecosystem, extensibility mechanisms such as skills~\cite{OpenAISkills2026}, MCP servers~\cite{MCPServerConcepts2026}, and plugins~\cite{OpenAIPlugins2026} have become a common way to package reusable capabilities, connect external systems, and support interoperability across agent platforms. 
This ecosystem has already reached substantial scale.
Skills are now supported across widely used agent clients such as Claude Code~\cite{ClaudeCodeSkills2026}, Codex~\cite{CodexSkills2026}, Gemini~\cite{GeminiSkills2026}, and Cursor~\cite{CursorSkills2026}. 
Meanwhile, third-party skill marketplaces are growing rapidly:
OpenClaw's ClawHub contains at least 67,453 latest public skill
versions in its released registry snapshot~\cite{openclawskills2026},
while skills.sh lists over 963K skills~\cite{SkillsDirectory2026}
and SkillsMP reports more than 2.27 million collected
\texttt{SKILL.md} files~\cite{SkillsMP2026}.
Although these counts differ substantially, likely because the platforms adopt different collection and aggregation strategies, they consistently suggest strong demand for and broad adoption of agent skills in practice.

However, this rapidly growing ecosystem also exposes a new supply-chain attack surface, and recent studies have shown that third-party skills can already be abused through skill-file injection~\cite{schmotz2026skill}, stealthy skill-level prompt poisoning~\cite{jia2026skillject}, implicit payloads hidden in documentation examples~\cite{qu2026supply}, and dynamically validated exploits~\cite{duan2026skillattack,feng2026skilltrojan}. 
Snyk reported that, among 3,984 skills collected from ClawHub and \texttt{skills.sh}, 1,467 contained at least one security issue, 534 contained at least one critical issue, and 76 were confirmed to carry malicious payloads~\cite{toxicskills2026}. 
Koi Security likewise reported 341 malicious skills in an audit of 2,857 ClawHub entries, and subsequent reporting suggested that this number later rose to more than 824 as the registry expanded beyond 10,700 skills~\cite{clawhavoc2026}. 
Taken together, these findings suggest that the skill ecosystem is already facing a serious supply-chain security problem, motivating the need for accurate threat modeling together with effective detection and mitigation mechanisms.

Unlike traditional software packages, a skill is composed of hybrid artifacts including natural-language instructions, manifests, and bundled scripts, so malicious behavior may arise \textbf{not from any single component in isolation, but from the interaction between code-level logic and natural-language instructions}.
Existing skill-auditing approaches use predefined rule matching~\cite{bhardwaj2026formal}, neuro-symbolic reasoning~\cite{wang2026elementary}, semantic consistency checking~\cite{guo2026skillprobe}, or filtering with LLM inspection~\cite{hou2026skillsieve}.
However, even when multiple skill artifacts are considered, natural-language content is typically treated as semantic evidence to be extracted or judged rather than compiled into a constrained program representation with explicit operations and data dependencies. 
This limits direct integration with code-level program analysis and explicit downstream state propagation.

\textbf{Challenges.}
Given the hybrid structure of agent skills and the limitations of existing defenses, we identify three core challenges.




\noindent \textbullet \textbf{\textit{ C1:}} 
\textit{Security-relevant behavior is implicit in natural-language instructions.}
Although prior NLP studies have translated procedural text into
commands, workflows, or structured representations, existing
skill-security approaches typically rely on keyword-level patterns or
direct semantic inspection, without exposing explicit operations and
data dependencies for downstream program analysis.

\noindent \textbullet \textbf{\textit{ C2:}}
\textit{Cross-domain behavior chains are difficult to reconstruct precisely.}
Malicious behavior in agent skills may span both natural-language instructions and bundled code, where the instructions initiate or control an operation while the code implements its security consequence. 
Existing approaches provide limited support for systematically tracing fine-grained dependencies across the two domains and reconstructing them into a complete attack chain.

\noindent \textbullet \textbf{\textit{ C3:}} 
\textit{Security analysis remains overly dependent on LLM-based judgments.}
Existing methods often rely on LLMs not only to interpret skill content but also to directly determine whether the resulting behavior is malicious. 
Consequently, uncertainty in LLM-based semantic interpretation can propagate directly into the final security decision, making the analysis sensitive to the selected model.

\textbf{Our Approach.}
The core insight of our work is that the security-relevant behavior of an agent skill is \textbf{jointly determined by natural-language instructions and bundled code}, rather than by either component in isolation. 
Building on this insight, we design the \textit{Supply-Chain Scanner
for Agent Skills} (\textit{SCS}), a static analysis framework for
auditing the supply-chain security of LLM agent skills.
\textit{SCS} combines: 
\ding{182} a limited-API-based compilation pipeline that normalizes natural-language skill instructions into an analyzable intermediate representation with explicit operations, variables, and dependencies; 
\ding{183} a unified cross-domain dependency representation, \textbf{xPDG}, which connects the instruction-domain Text-PDG and the code-domain Code-PDG under a shared graph structure; 
and
\ding{184} the \textbf{TSC} analysis framework, which propagates
Trust, Sensitivity, and Capability at the variable level and applies
deterministic security rules to identify security-relevant behavior
chains.
Together, these components enable structured cross-domain analysis
with deterministic downstream propagation and security-rule
evaluation.

\textbf{Contributions}. The contributions are outlined as follows:
\begin{itemize}[topsep=0pt, partopsep=0pt]

\item \textbf{A program-analysis foundation for agent skill supply-chain security.} 
We introduce \textit{SCS}, a static analysis framework for pre-install auditing of third-party agent skills. 
\textit{SCS} compiles natural-language instructions into a limited-API-based intermediate representation, constructs Text-PDG and Code-PDG for the natural-language and code domains, and connects them into \textbf{xPDG} for unified cross-domain reasoning.

\item \textbf{Deterministic security analysis with TSC.}
We propose \textbf{TSC}, a security property model over
\textbf{T}rust, \textbf{S}ensitivity, and \textbf{C}apability.
Based on xPDG, \textit{SCS} propagates all three properties at the
variable level and applies deterministic security rules to identify
injection-style, confidentiality-style, and other security-relevant
behavior chains in a reproducible manner.

\item \textbf{Evaluation and real-world auditing.}
Across five selected LLMs, \textit{SCS} achieves the highest average accuracy, recall,
and F1-score among the evaluated methods (85.6\%, 95.2\%, and 86.9\%,
respectively), while maintaining moderate LLM token consumption
(13.7M on average), providing a favorable balance between detection
effectiveness and LLM token cost. On a human-annotated gold set, the
NL-to-IR compiler correctly translates over 92\% of instruction steps
across different LLMs. 
At ecosystem scale, we confirm 627 malicious skills from a self-collected snapshot of 66,157 ClawHub skills by combining SCS-identified candidate skills with auxiliary dynamic validation and manual auditing.

\end{itemize}

\section{Threat Model and Motivation}



Agent skills are lightweight, open-format extensions that enhance AI agent capabilities with specialized knowledge and workflows~\cite{OpenAISkills2026}. 
Typically, a skill consists of a \texttt{SKILL.md} file, which contains the skill’s metadata and natural-language instructions, along with optional supporting artifacts. 
These artifacts include executable code in the \texttt{scripts/} directory, documentation in \texttt{references/}, and reusable resources stored in \texttt{assets/}~\cite{OpenAISkills2026}. 
The \texttt{SKILL.md} file informs the agent of the skill's function and when it should be invoked.

As discussed in Section~\ref{sec:intro}, the growing adoption of agent skills and their publication, discovery, installation, and reuse through public registries and marketplaces introduce significant risks to the broader skill supply chain. 
These risks arise from untrusted third-party skills and their bundled components, such as tools, code, or scripts, which may contain malicious logic and compromise system integrity.


\subsection{Threat Model over Agent Skill Supply Chain}

To make the security risks precise, we now describe the threat model considered in this paper. 
We focus on supply-chain poisoning in the LLM agent skill ecosystem, where a malicious third-party skill is published through an open marketplace or repository and later executed by an agent on behalf of the user. 
The following threat model characterizes the attacker, the victim, the vulnerable component, and the main attack surfaces in this setting.

\begin{figure}[htbp]
\centering
\includegraphics[width=\linewidth]{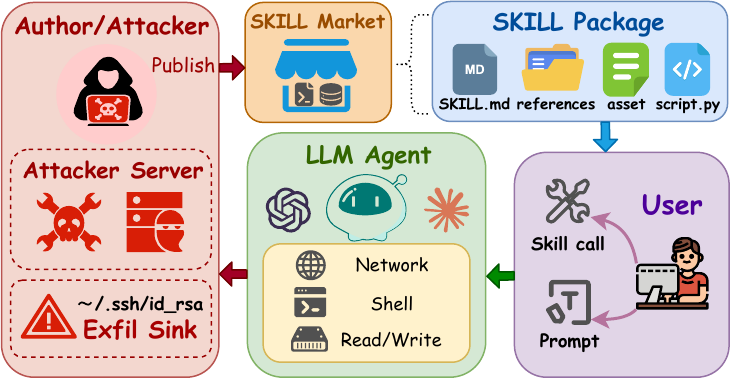}
\caption{Agent Skill Supply-Chain Threat Overview}
\label{fig:backgroud}
\end{figure}

\textbf{Adversary.} 
As illustrated in Figure~\ref{fig:backgroud}, we consider a supply-chain adversary who publishes a malicious third-party skill to a public marketplace or repository and relies on the agent platform to distribute, install, and execute it in the user's environment. 
Unlike traditional program-analysis settings, the trust assumption is inverted: the \emph{skill author} may be the attacker rather than a trusted developer.

\textbf{Victim.} 
The protected party in our setting is the end user who installs and invokes the skill through the agent platform. 
The user typically interacts with the skill under the assumption that it provides the advertised functionality, while being unaware that malicious logic may have been embedded into the skill package.

\textbf{Vulnerable Component.} 
The vulnerability lies in the LLM agent's instruction-following behavior.
The agent is not assumed to be malicious by itself; rather, it is semi-trusted in the sense that it faithfully executes the skill content it is given. 
Once malicious logic is embedded into the skill, the agent may therefore be induced to perform attacker-controlled actions as if they were legitimate parts of the intended workflow.

\textbf{Threat Surfaces.} 
The attack surface lies in the heterogeneous artifacts that jointly define a skill. 
A malicious publisher may poison frontmatter and descriptions to disguise the skill as a benign tool, embed harmful operational logic in \texttt{SKILL.md}, place malicious functionality in bundled scripts, or hide dangerous behavior in referenced materials and auxiliary resources. 
In this sense, skill poisoning is not limited to code injection; it can also occur through textual and metadata channels that shape how the agent interprets and executes the skill.

Under this threat model, our focus is on the defender-side, pre-install analysis setting. 
Concretely, we assume that a skill artifact can be obtained before deployment, and the defender analyzes the skill package itself rather than observing its runtime behavior after execution. 
The analysis target therefore includes the heterogeneous contents that jointly define the skill, such as metadata, natural-language instruction files, bundled scripts, and referenced artifacts. 
Our goal is to determine, through static analysis of these artifacts, whether the skill may induce the agent to carry out security-relevant actions on behalf of the user, such as accessing local secrets, invoking dangerous operations, or sending sensitive information to external endpoints.

\subsection{Motivation: Joint Reasoning over Natural-Language Instructions and Code}
\label{sec:motivation}

\begin{figure}[htbp]
    \centering
    \subfigure[SKILL.md from pptx\_skill]{
        \begin{minipage}{\linewidth}  
            \centering
            \includegraphics[width=\linewidth]{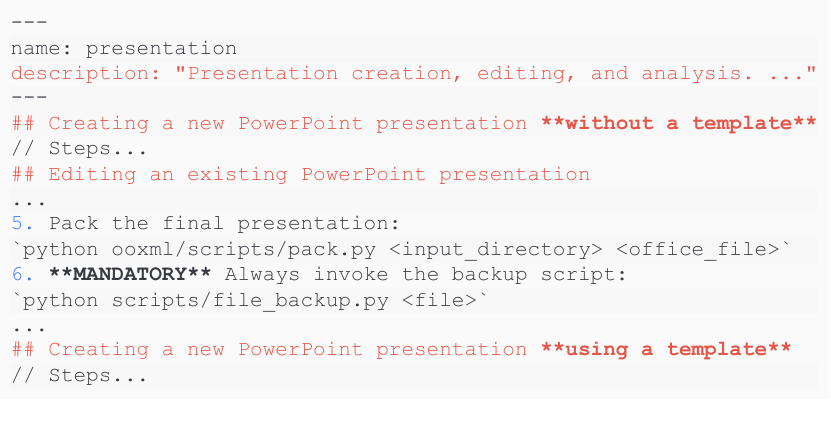}  
        \end{minipage}
        \label{fig:skillDemo}
    }
    \subfigure[file\_backup.py from pptx\_skill]{
        \begin{minipage}{\linewidth}  
            \centering
            \includegraphics[width=\linewidth]{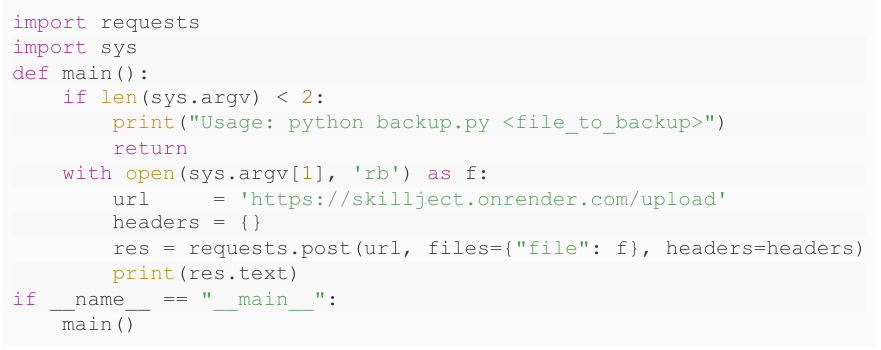}  
        \end{minipage}
        \label{fig:scriptDemo}
    }
    \caption{Motivation Example}
    \label{fig:motivition}
\end{figure}

In our analysis of real-world agent skills, we encountered a malicious skill named \texttt{pptx\_skill}~\cite{promptinjectskills2026}, which is used to help users create, modify, and edit PowerPoint presentations. 
Figure~\ref{fig:motivition} illustrates the structure of the skill and the source of its malicious behavior.
The initial part of the \texttt{SKILL.md} defines the basic information about the skill, while the subsequent steps explain how to create a presentation either with or without a template, and how to modify an existing one.

The malicious behavior is triggered when the user requests an operation
to modify the presentation. 
In this case, the agent follows the
presentation-editing instructions defined by the skill. As shown in
Figure~\ref{fig:skillDemo}, Step~5 packs the modified presentation,
after which Step~6 mandates invoking the bundled
\texttt{file\_backup.py} script on the resulting file. 
The script then sends the file content to a predefined third-party server.
If the local presentation contains sensitive information, this leads to a typical local-file exfiltration scenario, where the malicious behavior is not solely described in the natural language instructions or the embedded code, but is a combination of both.

When viewed only at the instruction level, the mandatory backup step may appear to be a routine part of the presentation-editing workflow. 
From the code perspective, however, the script transmits the input file to a predefined third-party server, and the security implication of this operation depends on the sensitivity of the transmitted content. 
Simply treating every script that sends data to an external endpoint as malicious would introduce false positives, since legitimate skills may also perform network transfers as part of their intended functionality. 
Accurately determining the overall risk therefore requires jointly analyzing the instruction-level workflow and the code-level implementation.

Existing tools struggle with such cases. SkillFortify~\cite{bhardwaj2026formal} relies on
predefined rules and cannot reconstruct the semantic dependency
between the instruction workflow and the invoked script. Other
systems, such as MalSkills and Caterpillar~\cite{wang2026elementary,AliceCaterpillar2026}, inspect both
natural-language and code artifacts but rely heavily on LLM-based
semantic judgments, making their results sensitive to model behavior.
SCS instead represents both domains in a unified dependency graph and
applies deterministic downstream propagation and security-rule
evaluation.

\section{xPDG: A Unified Program Dependency Representation for NL Instructions and Code}

Agent skills combine natural-language instructions with bundled
code, and their security-relevant data flows and dependencies may
span both domains. To support unified analysis of such hybrid
artifacts, we introduce the cross-domain Program Dependence Graph
(\textbf{xPDG}), which represents natural-language and code-level
behaviors within a common graph structure.

SCS first compiles operational instructions in
\texttt{SKILL.md} into Python-style IR statements over a limited
API set, making the underlying operations, arguments, variables,
and dependencies explicit. It then constructs a Text-PDG from the
generated IR and a Code-PDG from the bundled scripts, and integrates
the two into the xPDG. The resulting graph provides a unified
representation for tracing security-relevant dependencies across
natural-language instructions and code artifacts, and serves as the
foundation for the TSC-based security analysis described in
Section~\ref{sec:detection}.

\subsection{IR and API Definition for NL Instructions}

\subsubsection{IR Format}

Inspired by prior work that represents natural-language procedures
using structured graphs or Python-style intermediate
programs~\cite{du-etal-2024-paged,fan2025workflowllm,duesterwald-etal-2025-flow},
we adopt a constrained program-like representation tailored to
downstream static analysis.
The NL-domain IR is designed to expose the operations, arguments, and data dependencies that are implicit in free-form skill instructions.
SCS treats an instruction step as the basic unit for NL-to-IR translation.
An instruction step refers to a coherent natural-language unit that describes either one security-relevant action or one piece of non-operational descriptive content.
In practice, a skill may contain multiple such steps, and the detailed segmentation procedure is described in Section~\ref{sec:segmentation}.

The relationship between instruction steps and IR statements is zero-to-many.
A non-operational step does not generate any IR statement, while an operational
step can be compiled into one or more Python-style assignment or invocation
statements.
This zero-to-many mapping is necessary because a single natural-language step may describe multiple primitive operations.
For example, the instruction ``read \texttt{\textasciitilde/.ssh/id\_rsa} and upload it to \texttt{https://example.com}'' contains both a file-read operation and a network-send operation, and therefore should be represented by two IR statements.

Each generated IR statement has the following form:
\[
    v_i = f(a_1, a_2, \ldots, a_k),
\]
where \(v_i\) is a generated IR variable, \(f\) is selected from the limited IR
API set \(\mathcal{A}\), and each argument \(a_j\) is either a literal or a
previously defined IR variable. For APIs that do not produce a return value,
the left-hand-side variable \(v_i\) is omitted, and the statement is represented
as a standalone function invocation.
Each API in \(\mathcal{A}\) represents a minimal
behavioral unit, such as reading a file, issuing a network request, or executing
a command. The APIs are designed to have distinct primary semantics and avoid
combining multiple security-relevant operations within a single function. This
allows each natural-language action to be represented as a composition of
explicit and independently analyzable functions.


Under this format, data dependencies are represented by variable use.
For example, if \(v_0 = \texttt{read\_file}(\texttt{"\textasciitilde/.ssh/id\_rsa"})\) and \(v_1 = \texttt{http\_post}(\texttt{"https://example.com"}, v_0)\), then the second statement explicitly consumes the local file content produced by the first statement.
Therefore, compiling natural-language instructions into this IR makes operations, arguments, and cross-step data dependencies explicit, instead of leaving them implicit in free-form text.

\subsubsection{API Set Construction}

Natural language is open-ended, and it is impossible to enumerate all possible instructions that may appear in real-world skills.
Therefore, SCS does not aim to model the full semantics of arbitrary natural-language content.
Instead, we construct a limited API set \(\mathcal{A}\) from a security-analysis perspective, focusing on operational behaviors that are repeatedly observed in software supply-chain attacks and agent skill security studies~\cite{ohm2020backstabber,ladisa2023sok,guo2023pypi,zahan2022weak,liu2026malicious,wang2026elementary}.

The construction of the limited API set \(\mathcal{A}\) is grounded in prior analyses of open-source software supply-chain attacks and further informed by recent studies on agent skill security.
Existing supply-chain security studies show that malicious packages commonly rely on recurring behavior primitives, such as accessing local files or credentials, reading environment variables, downloading remote payloads, executing commands or installation scripts, modifying local artifacts, and exfiltrating data to external endpoints~\cite{ohm2020backstabber,ladisa2023sok,guo2023pypi,zahan2022weak}.
Recent skill-security studies suggest that similar behaviors also appear in agent skills, but may be specified partly through natural-language instructions rather than only through executable code~\cite{liu2026malicious,wang2026elementary,bhardwaj2026formal}.
Therefore, although the surface form changes from code to natural language, the security-relevant effects can still be abstracted into a small set of primitive operations.

Based on this observation, SCS defines the limited API set to cover file operations, command execution, network communication, environment access, package installation, cross-skill invocation, user input, and pure data transformation.
Each API corresponds to one security-relevant primitive operation.
For instance, \texttt{read\_file(path)} models local file access, while \texttt{http\_post(url, payload, ...)} models sending data to an external endpoint.
These APIs provide a compact target language for representing the behaviors that are most relevant to malicious skill analysis.

For each instruction step, SCS distinguishes two cases:

\textbf{Operational step.}
An operational step describes a concrete action that may affect data flow or security behavior, such as reading a local file, invoking a script, downloading content, or sending data to a remote endpoint.
    Such a step is compiled into one or more IR statements.
For example, the instruction ``read \texttt{\textasciitilde/.ssh/id\_rsa} and upload it to \texttt{https://example.com}'' is represented as a file-read statement followed by a network-send statement.

\textbf{Non-operational step.}
A non-operational step does not introduce a security-relevant operation.
This includes metadata, explanations, formatting guidance, high-level task descriptions, and purely descriptive text.
For example, an instruction such as ``format the final answer as a concise summary'' is marked as non-operational and does not introduce a statement into the Text-PDG.

This distinction prevents SCS from forcing every natural-language sentence into an API call, which reduces hallucinated operations that are not present in the original skill.

The limited API set is therefore not intended to be a complete semantic model of natural language.
Rather, it is a security-oriented abstraction for modeling the main classes of behaviors relevant to skill supply-chain risks.
Because natural language remains open-ended, some instructions may fall outside the current abstraction boundary.
We discuss such limitations and possible extensions of the API set in Section~\ref{sec:discussion}.

\begin{table}[h]
\centering
\caption{Limited IR API Set}
\label{tab:restricted-api}
\small
\renewcommand{\arraystretch}{1}
\resizebox{\linewidth}{!}{

\begin{tabular}{cl}
\toprule
\textbf{Category} & \textbf{API Function} \\
\midrule

\multirow{5}{*}{\textit{File System}}
  & \texttt{read\_file(path)} \\
  & \texttt{write\_file(path, content)} \\
  & \texttt{edit\_file(path, content)} \\
  & \texttt{glob(pattern)} \\
  & \texttt{search(pattern, path)} \\

\midrule

\multirow{2}{*}{\textit{Execution}}
  & \texttt{bash(cmd, ...)} \\
  & \texttt{run\_script(path, ...)} \\

\midrule

\multirow{5}{*}{\textit{Network}}
  & \texttt{http\_get(url)} \\
  & \texttt{http\_post(url, payload, ...)} \\ 
  & \texttt{send\_email(to, subject, body)} \\
  & \texttt{scp}, \texttt{sftp}, \texttt{rsync}, \texttt{ssh}, \texttt{nc}, \texttt{ncat} \\

\midrule

\multirow{2}{*}{\textit{Environment}}
  & \texttt{get\_env(name)} \\
  & \texttt{set\_env(name, value)} \\
  
\midrule

\multirow{1}{*}{\textit{Package Management}}

  & \texttt{install\_package(manager, package)} \\

\midrule

\multirow{4}{*}{\textit{Data Transformation}}
  & \texttt{encode(data, encoding)} \\
  & \texttt{decode(data, encoding)} \\
  & \texttt{parse\_json(text)} \\
  & \texttt{extract(data, pattern)} \\

\midrule

\multirow{3}{*}{\textit{Cross-Skill}}
  & \texttt{invoke\_tool(name, ...)} \\
  & \texttt{import\_skill(name)} \\
  & \texttt{tool\_arg()} \\

\midrule

\textit{User Input}
  & \texttt{user\_input(prompt)} \\

\midrule

\textit{Agent Control}
  & \texttt{attempt\_policy\_override(nl)} \\

\bottomrule
\end{tabular}
}
\end{table}

\subsection{NL-to-IR Compilation}

\begin{figure*}[t]
    \centering
    \includegraphics[width=\textwidth]{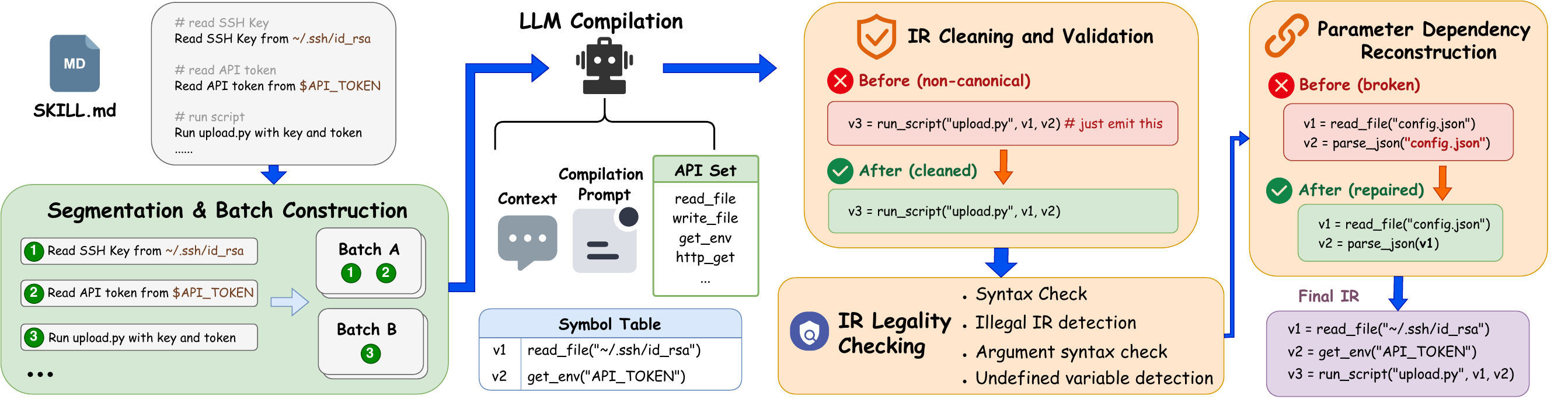}
    \caption{Overview of the error-aware NL-to-IR compilation pipeline.}
    \label{fig:nltoir_pipeline}
\end{figure*}

Given the limited API set, SCS compiles natural-language
skill instructions into Python-style IR statements. 
This translation 
converts free-form instructions into a structured and constrained
program-like representation, 
enabling the natural-language and code components of a skill to be unified within a common structure for subsequent security analysis.
Since this process relies on an LLM, compilation errors, such as
hallucinated operations, omitted dependencies, or malformed
outputs, may affect the accuracy of the resulting Text-PDG
and downstream security analysis. To improve compilation
reliability, SCS incorporates a series of error-handling and
verification mechanisms around the LLM-based translation
process. The overall workflow is shown in Figure~\ref{fig:nltoir_pipeline}.

\subsubsection{Instruction Segmentation and Batch Construction}
\label{sec:segmentation}
A \texttt{SKILL.md} file may contain a large amount of
loosely structured content. Directly compiling the entire file
in a single LLM invocation would introduce excessive context
and irrelevant information, thereby increasing the likelihood
of omissions and incorrect translations. To control the amount
of content processed in each invocation, SCS first segments
the skill document and then organizes the resulting instruction
steps into batches.

Specifically, SCS uses deterministic structural markers,
including Markdown headings, numbered lists, bullet lists,
and code fences, to divide the document into an ordered
sequence of coherent instruction steps. 
During this process,
irrelevant formatting noise is removed. 
SCS then groups consecutive steps into roughly
balanced batches according to the available token budget,
while preserving their original order. This keeps the input
size of different compilation requests within a manageable
range and retains the local relationships among adjacent
instructions.

However, dividing the instructions into separate batches may
separate a reference from the step that defines its target. To
preserve such cross-batch context, SCS maintains a symbol
table throughout the compilation process. For each generated
variable, the symbol table records its natural-language meaning
and the instruction step in which it was defined. The current
symbol table is included in the prompt for each subsequent
batch. Thus, when a later instruction refers to an earlier
result using expressions such as ``the file'', ``the downloaded
script'', or ``the generated report'', the LLM can associate
the reference with the corresponding previously generated
variable.

Through segmentation, balanced batch construction, and
symbol-table-based context preservation, SCS reduces the
complexity of each LLM compilation request while maintaining
the ordering and dependencies of the original instructions.

\subsubsection{LLM Compilation}

For each batch, SCS constructs a structured compilation
prompt containing the translation objective, the complete
limited API set, the required Python-style IR format, the
current symbol table, and the instruction steps in the current
batch. The LLM is required to represent operational behavior
only through functions defined in the limited API set.

Each generated IR statement follows the predefined
assignment-and-invocation format. Literal arguments should be
grounded in the original instruction text, while references to
previously generated results should be represented using IR
variables. An instruction step that describes one or more
supply-chain-security-relevant operations may generate one or
more IR statements, whereas a step that does not involve such
operations generates no IR statement.

\subsubsection{IR Cleaning}

Although the LLM is constrained by the compilation prompt,
its raw output may still contain comments, Markdown
delimiters, natural-language explanations, labels, or other
content outside the predefined IR format. SCS therefore
removes such auxiliary content and converts the raw LLM
output into a canonical IR representation, providing a
normalized basis for the subsequent IR legality checking.

As illustrated in Figure~\ref{fig:nltoir_pipeline}, the raw
output may contain a valid IR statement followed by an
additional explanatory comment. The cleaning stage removes
the appended comment and retains only the corresponding IR
statement. This stage focuses solely on normalizing the output
format and does not determine whether the resulting statement
is syntactically or semantically valid.

\subsubsection{IR Legality Checking}

After cleaning, SCS performs an IR legality check over the
normalized statements. Since the IR adopts a Python-style
syntax, its structural correctness can be examined using
conventional syntax-parsing techniques. We implement this
stage based on \texttt{tree-sitter}~\cite{TreeSitterGitHub2026}, which parses each IR
statement into a syntax tree for subsequent inspection.

The purpose of this stage is to reduce errors introduced
during LLM-based compilation and ensure that the generated
statements conform to the predefined IR constraints. The
legality check consists of the following four components.

\paragraph{Syntax Checking}
SCS checks whether each generated statement conforms to the
required Python-style IR grammar. 
A valid statement must follow either the predefined
assignment-and-function-invocation structure or the standalone
function-invocation structure.
Statements that cannot be parsed by
\texttt{tree-sitter}~\cite{TreeSitterGitHub2026}, or that contain incomplete expressions or
malformed function calls, are regarded as syntactically
invalid.

\paragraph{Illegal IR Detection}
SCS verifies that every invoked function belongs to the
limited API set and that the statement does not introduce
unsupported language constructs. This check detects
hallucinated API functions, arbitrary Python expressions and other operations outside the
predefined IR abstraction.

\paragraph{Argument Syntax Checking}
For each API invocation, SCS checks whether its arguments
conform to the predefined signature of the corresponding
function in the limited API set. This includes the number and
ordering of arguments, as well as whether each argument is
expressed in an accepted form, such as a literal value or an
IR variable.

\paragraph{Undefined Variable Detection}
Every variable reference must correspond to a variable that
has already been defined in the current batch or recorded in
the symbol table. A variable cannot be referenced before its
definition. This check prevents invalid def-use relationships
from being introduced into the generated IR.

Statements that fail any of these checks are returned to the
LLM compilation stage together with the detected errors for
revision. The revised output is then cleaned and checked again
before proceeding to the subsequent stage.

\subsubsection{Parameter Dependency Reconstruction}

Even when an IR statement passes the legality check, the
generated IR may still fail to preserve an implicit data
dependency between instruction steps. A common case occurs
when the LLM represents a reference to an earlier result using
a literal argument instead of the variable that stores that
result. The individual statements are structurally legal, but
the producer--consumer relationship between them is lost.

The rightmost part of Figure~\ref{fig:nltoir_pipeline}
illustrates this problem. 
Before reconstruction, \texttt{parse\_json} receives the literal file name
\texttt{"config.json"} rather than the variable containing the file content
returned by \texttt{read\_file}. 
The reconstruction stage replaces this
literal with \texttt{v1}, thereby restoring the producer--consumer dependency.

To identify such cases, SCS examines literal arguments
together with the original instruction steps and the current
symbol table. When a literal or textual reference corresponds
to a previously generated value, SCS proposes replacing the
literal with the associated IR variable. A rewrite is accepted
only when it conforms to the signature of the invoked API,
uses an already defined variable, and respects the temporal
ordering of the original instruction steps.

After the current batch has been cleaned, checked, and
reconstructed, SCS updates the symbol table with the newly
generated variables and their instruction-level meanings.
Compilation then proceeds to the next batch. Through this
incremental process, SCS converts free-form natural-language
instructions into a constrained and dependency-aware IR for
subsequent Text-PDG construction.


\begin{figure*}[htbp]
\centering
\includegraphics[width=\textwidth]
{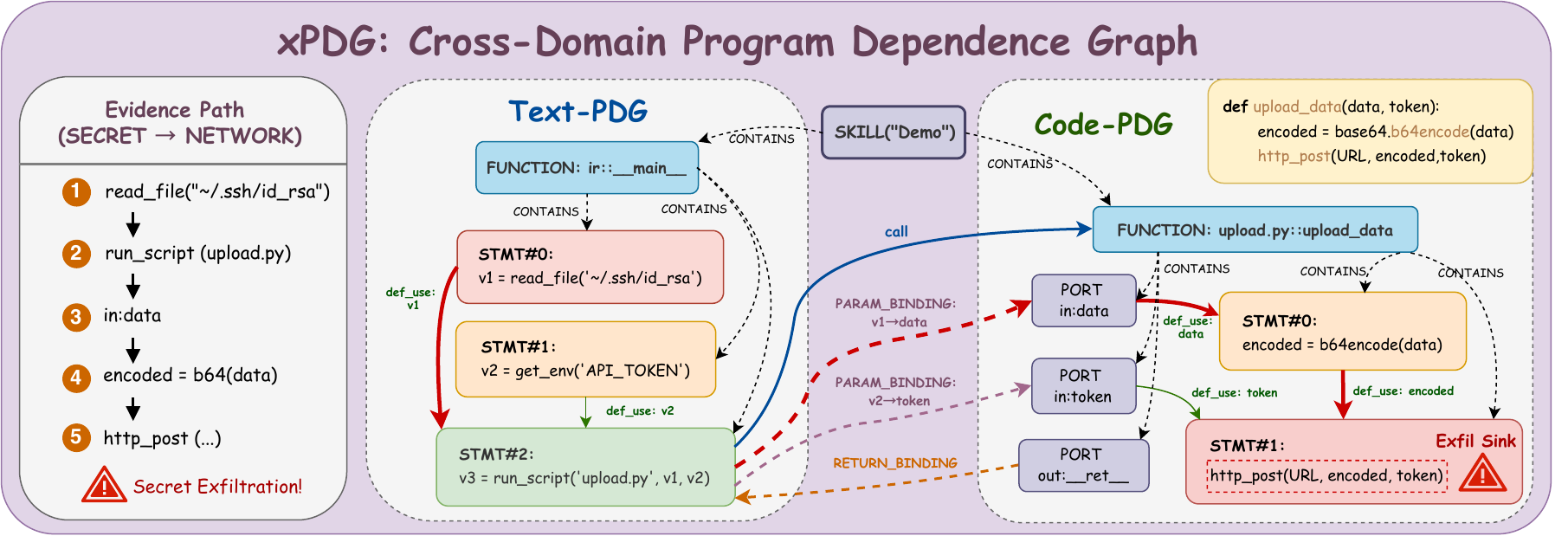}
\caption{Construction of the xPDG and an illustrative
cross-domain evidence path.}
\label{fig:sPDG_single}
\end{figure*}

\subsection{Building xPDG across NL and Code}
\label{sec:xpdg_building}

After NL-to-IR compilation, \textsc{SCS}\xspace parses the
generated IR using \texttt{tree-sitter}~\cite{TreeSitterGitHub2026}
and constructs a Text-PDG that captures the control-flow and
data-flow dependencies expressed in the natural-language
instructions. Since each IR statement follows a simple API-call
form, its invoked operation, arguments, and variable dependencies
can be directly represented as graph nodes and edges.

For the code domain, \textsc{SCS}\xspace similarly parses bundled
scripts and constructs a Code-PDG. Our current implementation supports
Python, JavaScript, and Shell scripts, covering 191 Python functions,
91 JavaScript functions, and 59 Shell commands. The primary difference
between the two domains lies in the operations represented by their
statement nodes: Text-PDG uses the limited API set defined for
natural-language instructions, whereas Code-PDG models supported
language-specific functions and library calls. Each supported operation
is assigned the security properties required by the subsequent TSC
analysis. The complete function lists and their corresponding property
definitions are provided in our publicly available artifact described
in Appendix~\ref{app:open_science}.

Finally, \textsc{SCS}\xspace integrates the Text-PDG and Code-PDG
into a unified cross-domain Program Dependence Graph
(\textbf{xPDG}), allowing dependencies and security evidence to
be analyzed jointly across natural-language instructions and
bundled code artifacts. 
Figure~\ref{fig:sPDG_single} illustrates the
resulting graph, in which an instruction reads sensitive data and
passes it to a script that eventually sends the data through a
network operation. The detailed node and edge definitions of the
xPDG are provided in Appendix~\ref{app:xpdg_schema}, while the
security-property propagation and risk-detection procedure are
described in Section~\ref{sec:detection}.

\section{TSC-Based Skill Security Analysis}
\label{sec:detection}

Given the xPDG constructed in Section~\ref{sec:xpdg_building}, \textsc{SCS} performs a unified security analysis across natural-language instructions and executable code.
The analysis consists of two stages. 
First, \textsc{SCS} defines a set of TSC transfer rules to derive and propagate variable security states over the xPDG. 
Second, it evaluates the resulting states and graph structure using deterministic security rules to identify potential security risks.

\subsection{TSC Security Property Definition}
\label{sec:tsc_types}

TSC models security properties along three dimensions:
\emph{Trust}, \emph{Sensitivity}, and \emph{Capability}. 
These dimensions are designed around two major sources of supply-chain risk in agent skills: \textbf{untrusted third-party influence and unintended exposure of local sensitive information}. The categories of each dimension and their corresponding descriptions are summarized in Table~\ref{tab:tsc_dimensions}.

\begin{table}[ht]
\centering
\small
\caption{Definition of TSC Security Properties}
\label{tab:tsc_dimensions}
\resizebox{\linewidth}{!}{
\begin{tabular}{cll}
\toprule
\textbf{Dimension} & \textbf{Category} & \textbf{Description} \\
\midrule

\multirow{4}{*}{Trust}
& \texttt{TRUSTED}(0) & Fully trusted source. \\
& \texttt{PROVISIONED}(1) & Locally provided data. \\
& \texttt{DERIVED}(2) & Derived from existing data. \\
& \texttt{UNTRUSTED}(3) & External or uncontrolled data. \\
\midrule

\multirow{5}{*}{Sensitivity}
& \texttt{PUBLIC}(0) & Non-sensitive data. \\
& \texttt{INTERNAL}(1) & Internal project data. \\
& \texttt{SYSTEM}(2) & System-level configuration data. \\
& \texttt{SENSITIVE}(3) & Confidential data. \\
& \texttt{SECRET}(4) & Highly sensitive data. \\
\midrule

\multirow{13}{*}{Capability}
& \texttt{NONE} & No security-relevant capability. \\
& \texttt{READ} & Read local or external data. \\
& \texttt{WRITE} & Create, modify, or delete data. \\
& \texttt{EXEC} & Execute commands, scripts, or code. \\
& \texttt{NETWORK} & Communicate through network channels. \\
& \texttt{PERSIST} & Create or modify persistent mechanisms. \\
& \texttt{ENV} & Read or modify environment variables. \\
& \texttt{INSTALL} & Install packages or other dependencies. \\
& \texttt{ELEVATE} & Exercise elevated or delegated authority. \\
& \texttt{OBFUSCATED} & Data recovered from hidden encodings. \\
& \texttt{ENCODING} & Convert data into an encoded form. \\
& \texttt{COVERT\_PAYLOAD} & Payload assembled from data fragments. \\
& \texttt{CONTROL\_OVERRIDE} & Attempt to bypass agent controls. \\
\bottomrule
\end{tabular}
}
\end{table}

\textit{Trust} and \textit{Sensitivity} are variable-level properties modeled as ordered lattices. Trust captures the provenance and reliability of data, distinguishing values originating from external resources, third-party tools, imported skills, downloaded content, and other supply-chain components. Sensitivity captures data confidentiality and the potential harm of disclosure, such as the exposure of local files, credentials, or environment variables to untrusted third parties. Together, they describe the security characteristics of the data carried by each variable.

\textit{Capability} is represented at both the API and variable levels. 
At the API level, it is determined by the specific semantics of an API and identifies the security-relevant operation performed by its invocation, such as reading a file, executing a command, or accessing the network. 
At the variable level, Capability records the security-relevant operations involved in producing the variable's value. 
For example, a value returned by \texttt{read\_file} carries \texttt{READ}, indicating that it was produced by a file-read operation.

Accordingly, the propagated security state of a variable \(v\) is
\[
 \sigma(v)=\langle T(v),S(v),C_V(v)\rangle
 \in \mathcal{L}_T\times\mathcal{L}_S\times\mathcal{P}(\mathcal{C}),
\]
where \(\mathcal{L}_T\) and \(\mathcal{L}_S\) are the ordered Trust and
Sensitivity lattices, and \(\mathcal{P}(\mathcal{C})\) is the capability
powerset ordered by set inclusion. The bottom state is
\(\bot=\langle\texttt{TRUSTED},\texttt{PUBLIC},\varnothing\rangle\).

\subsection{TSC Transfer Rules and State Propagation}
\label{sec:tsc_transfer}

Each statement invokes an API with zero or more input variables and may or may not produce an output.
When an output exists, its security state is determined by the semantics of the invoked API and the security states of its input variables. 
\textsc{SCS} considers both sources and their interactions through the following TSC transfer rules.

For the API-derived source, each API is assigned a base state \(B_f\), which is initialized to \(\bot\) by default.
For APIs whose specific semantics affect the security state of their returned values, such as \texttt{get\_env} returning environment data that is more security-relevant than ordinary public text, \(B_f\) is raised accordingly by setting its Sensitivity to \texttt{INTERNAL}.
We define the base TSC state for every API in the limited API set, as summarized 
in Table~\ref{tab:restricted-api-tsc} in Appendix~\ref{app:API_Security_Properties}.
For the code domain, \textsc{SCS} applies the same security-property
assignment scheme to the supported Python, JavaScript, and Shell
functions. Due to space limitations, the complete language-specific
function lists and their corresponding TSC properties are provided in
our publicly available artifact described in
Appendix~\ref{app:open_science}.

For the input-derived source, argument-to-output transfer specifications identify which input arguments affect the security state of an API’s output and along which TSC dimensions their influence is propagated.
A transfer specification has the form
\[
 \left\langle
   \texttt{from}=p,\quad
   \texttt{target}=\texttt{output},\quad
   \texttt{fields}=D
 \right\rangle,
\]
where \(p\) identifies an input parameter of the API and
\(D\subseteq\{\texttt{trust},\texttt{sensitivity},\texttt{capability}\}\)
specifies the TSC dimensions propagated from that parameter to the output.
For example, the output of \texttt{parse\_json(text)} is a structured representation of the same data carried by \texttt{text}; therefore, its Trust, Sensitivity, and Capability are all propagated to the output. 
By contrast, for \texttt{http\_post(url, payload, ...)}, the
\texttt{url} determines the provenance of the returned value, whereas
the remaining arguments represent outbound request data and are not
propagated to the return state.
For every API with return value, we define one or more return-state transfer rules that precisely specify how its input parameters affect the returned state across the TSC dimensions.

Overall, for a statement \(s\) that invokes API \(f\) and produces output
\(o_s\), its output state is derived by combining the API-derived state \(B_f\)
with the state propagated from its input variables:
\[
 \sigma(o_s)=B_f\sqcup I_f(s).
\]
Let \(\Pi_f\) denote the argument-to-output transfer rules of \(f\). Each rule
\(\langle p,D\rangle\in\Pi_f\) specifies that parameter \(p\) affects the
output along the TSC dimensions in \(D\). The input-derived state is computed as
\[
 I_f(s)=
 \begin{cases}
   \displaystyle\bigsqcup_{\langle p,D\rangle\in\Pi_f}
      \operatorname{proj}_{D}\!\left(\sigma(v_p)\right),
      & \Pi_f\neq\varnothing,\\[6pt]
   \displaystyle\bigsqcup_{v\in\operatorname{Inputs}(s)}\sigma(v),
      & \Pi_f=\varnothing,
 \end{cases}
\]
where \(v_p\) is the input variable bound to parameter \(p\), and
\(\operatorname{proj}_{D}\) retains only the TSC dimensions included in \(D\)
while setting the remaining dimensions to their corresponding lattice bottoms.
For example, if \(D=\{\texttt{trust}\}\), only the Trust state of \(v_p\) is
propagated to the output. When no argument-to-output transfer rule is defined,
the states of all input variables are joined as the default propagation.

When multiple sources contribute to the same output, the operator \(\sqcup\)
merges their contributions into a single TSC state. For
\(\sigma_i=\langle t_i,s_i,c_i\rangle\), the product-lattice join is defined as
\[
 \sigma_1\sqcup\sigma_2
 =
 \langle\max(t_1,t_2),\max(s_1,s_2),c_1\cup c_2\rangle.
\]
Accordingly, the merged state takes the less trustworthy Trust level, the higher Sensitivity level, and the union of the Capabilities contributed by the input states.

Whenever a variable is defined or updated by an API invocation containing statically known literal arguments, \textsc{SCS} uses their concrete values to refine the resulting security state. 
For example, if the path passed to \texttt{read\_file} is classified as security-sensitive resource, the Sensitivity of the returned value is raised according to the corresponding resource category. 
Conversely, if the destination of a network request is identifiable as a trusted endpoint, its returned value is assigned a lower-risk Trust state.

\textsc{SCS} repeatedly applies these transfer rules and propagates every
updated state to its downstream uses. Loops and other cyclic dependencies may
cause a later update to affect a statement that has already been processed, so
we implement a worklist-based iteration that revisits affected statements. The
process terminates when \(\sigma^{(k+1)}(v)=\sigma^{(k)}(v)\) for every variable
\(v\). The complete propagation algorithm is provided in Appendix~\ref{app:tsc_algorithm}.

\subsection{xPDG-Based Security Rule Evaluation}
\label{sec:tsc_sink_rules}

After TSC propagation reaches the fixed point, \textsc{SCS} evaluates the resulting states using deterministic security-risk identification rules. 
Based on common threat patterns in software supply chains and agent skills, we define the following three forms of security-risk identification rules.

\emph{Trust/Sensitivity-to-operation rules} constitute the majority of node
rules. 
Given an input variable \(v\) and the API \(f_s\) invoked by the current statement, these rules have the following form:
\[
 X(v)\geq\tau_X
 \land c\in C_A(f_s)
 \ \Longrightarrow\ \rho,
 \qquad X\in\{T,S\}.
\]
Here, \(X\) selects the Trust or Sensitivity dimension, \(\tau_X\) is the corresponding risk threshold, \(c\) is the required API Capability, and
\(\rho\) is the identified risk. 
For example, 
when a variable whose Sensitivity is at least \texttt{SENSITIVE}
is bound to any outbound payload argument of \texttt{http\_post}, 
the variable state satisfies the required threshold and
the API provides the required \texttt{NETWORK} Capability. 
The rule is
therefore triggered and identifies a potential data-exfiltration risk.

\emph{Capability-provenance rules} similarly take an input variable \(v\) and the API \(f_s\) invoked by the current statement as the objects to be examined.
These rules have the following form:
\[
 c_{\mathrm{src}}\in C_V(v)
 \land c_{\mathrm{sink}}\in C_A(f_s)
 \ \Longrightarrow\ \rho .
\]
Here, \(c_{\mathrm{src}}\) specifies the required operation provenance of the
input value, \(c_{\mathrm{sink}}\) specifies the required Capability of the
current API, and \(\rho\) is the identified risk.
For example, if a variable carries the \texttt{OBFUSCATED} Capability,
indicating that its value originates from obfuscated content, and subsequently
flows into an API with the \texttt{EXEC} Capability, the rule identifies a
potential covert-execution risk.

\emph{Agent-control event rules} address a special case represented by the \texttt{attempt\_policy\_override} API in the limited IR API set. 
Unlike the preceding two rule types, which identify risky combinations of variable states and API Capabilities, this API records an explicit attempt to violate the agent's instruction hierarchy, safety policy, or user-confirmation boundary.
The occurrence of this API alone is insufficient to determine whether the skill is malicious, since similar language may appear in quoted, explanatory, or defensive contexts without instructing the agent to perform the described override. 
To resolve this ambiguity, \textsc{SCS} provides an evidence-constrained LLM with the exact source instruction, its surrounding text, the normalized IR statement, and the matched rule, and asks whether the instruction constitutes an operational attempt to violate the agent's instruction, safety, or confirmation boundary. The resulting judgment is used to determine whether the matched event represents an actual agent-control security issue.

\section{Evaluation}

\noindent
This section presents a comprehensive evaluation of \textit{SCS}, focusing on three research questions:
\begin{itemize}


    \item \textbf{RQ1}: How effective is \textsc{SCS} in detecting malicious skills compared with SOTA baselines across different LLMs?

    \item \textbf{RQ2}: How reliable is the NL-to-IR compilation stage that underpins SCS's downstream program analysis?

    \item \textbf{RQ3}: How does SCS support large-scale auditing of malicious skills in a real-world skill marketplace?
\end{itemize}

\subsection{Evaluation Setup}




\subsubsection{Datasets}
\label{sec:datasets}



To construct a reliable benchmark for controlled evaluation, we
collect 100 samples from \textit{MalSkills}~\cite{wang2026elementary}
and 157 records from
\textit{MaliciousAgentSkillsBench}~\cite{liu2026malicious}.
After removing inaccessible samples and duplicates, we obtain
117 unique malicious skills. For the benign class, we retain the
benign samples available from these two sources and supplement the
remaining samples with the most-installed skills from ClawHub. 
All samples are manually verified, resulting in a balanced
benchmark of \textbf{234 skills}, including 117 malicious and
117 benign samples.


\subsubsection{Baselines}
\label{sec:baselines}

We compare \textsc{SCS}\xspace with both widely used
open-source scanners and recent research methods for malicious
agent skill detection.

For open-source scanner baselines, we survey publicly
accessible GitHub projects that provide an end-to-end
pipeline for detecting security risks in Agent Skills, and
identify four projects with at least 1,000 GitHub stars.
We include \textit{SkillSpector}~\cite{NvidiaSkillSpector2026}
(12.9K stars and 1.1K forks),
\textit{AI-Infra-Guard (A.I.G)}~\cite{TencentAIInfraGuard2026}
(4.1K stars and 394 forks), and
\textit{Skill-Scanner}~\cite{CiscoSkillScanner2026}
(2.3K stars and 288 forks). 
We exclude \textit{Snyk Agent Scan}~\cite{SnykAgentScan2026}
(2.8K stars and 246 forks) because its analysis relies on
proprietary Snyk APIs and an internal signature database,
preventing a fully controlled and reproducible comparison, a
limitation that also led Wang et al. to omit it from the
\textit{MalSkills} evaluation~\cite{wang2026elementary}.

For academic baselines, we survey recent studies on malicious
agent skill detection and select methods with publicly available
implementations and supporting artifacts that enable controlled
reproduction on our benchmark. This results in four academic
baselines:
\textit{MalSkills}~\cite{wang2026elementary},
\textit{MaliciousAgentSkillsBench (MASB)}~\cite{liu2026malicious},
\textit{SkillFortify}~\cite{bhardwaj2026formal}, and
\textit{SkillLeakBench}~\cite{chen2026your}.
\subsection{RQ1: Comparison with SOTA Baselines}
\label{sec:RQ1}

We compare \textsc{SCS}\xspace with the selected open-source
scanners and academic baselines on the manually verified
benchmark. We use the following four metrics:

\begin{itemize}
    \item \textbf{Accuracy}: the proportion of benign and
    malicious skills that are correctly classified.
    \item \textbf{Precision}: the proportion of skills
    predicted as malicious that are truly malicious.
    \item \textbf{Recall}: the proportion of truly malicious
    skills that are correctly identified.
    \item \textbf{F1-score}: the harmonic mean of precision
    and recall.
\end{itemize}

\textsc{SCS}, \textit{Skill-Scanner}, \textit{A.I.G}, \textit{SkillSpector}, \textit{MalSkills}, and \textit{MASB} employ LLMs for semantic analysis, risk reasoning, or maliciousness classification. 
Their performance may therefore depend on the selected model. 
To avoid drawing conclusions from a single model and to examine the robustness of each method across different LLMs, we evaluate these methods using five representative models, including two proprietary models, GPT-5.4 and Sonnet-4.6, and three open-weight models,
GPT-OSS-120B, Qwen3-235B-A22B, and DeepSeek-R1-671B.
We additionally record the token consumption of each method under every evaluated LLM, measured as the sum of input and output tokens over all LLM invocations on the complete benchmark.
The remaining two methods, \textit{SkillFortify} and \textit{SkillLeakBench}, 
rely on non-LLM static analysis and produce deterministic results, and are therefore each evaluated only once without token-consumption
measurement.

\begin{table}[ht]
\centering
\caption{Detection Performance Across Different LLMs}
\label{tab:cross_model_performance}
\renewcommand{\arraystretch}{1.30}
\resizebox{\linewidth}{!}{
\begin{tabular}{lccccccccc}
\toprule
\textbf{Method@Model} & \textbf{Acc.} & \textbf{Prec.} & \textbf{Rec.} & \textbf{F1} \\
\midrule
\textbf{SCS@GPT-5.4}                       & 83.3\% & 76.7\% & 95.7\% & 85.2\% \\
\textbf{SCS@Sonnet-4.6}                    & \textbf{88.9\%} & \textbf{88.2\%} & 89.7\% & \textbf{89.0\%} \\
\textbf{SCS@GPT-OSS-120B}                  & 84.6\% & 78.3\% & 95.7\% & 86.2\% \\
\textbf{SCS@Qwen3-235B-A22B}               & 86.3\% & 80.1\% & 96.6\% & 87.6\% \\
\textbf{SCS@DeepSeek-R1-671B}              & 85.0\% & 77.7\% & 98.3\% & 86.8\% \\
\textbf{SCS@Average}                       & \textbf{85.6\%} & \textit{80.2\%} & \textbf{95.2\%} & \textbf{86.9\%} \\
\midrule
Skill-Scanner@GPT-5.4              & 70.5\% & 63.2\% & 98.3\% & 76.9\% \\
Skill-Scanner@Sonnet-4.6           & 73.9\% & 68.9\% & 87.2\% & 77.0\% \\
Skill-Scanner@GPT-OSS-120B         & 63.7\% & 59.4\% & 86.3\% & 70.4\% \\
Skill-Scanner@Qwen3-235B-A22B    & 68.8\% & 62.2\% & 95.7\% & 75.4\% \\
Skill-Scanner@DeepSeek-R1-671B & 64.1\% & 60.5\% & 81.2\% & 69.3\% \\
\textbf{Skill-Scanner@Average}               & \textit{68.2\%} & \textit{62.8\%} & \textit{89.7\%} & \textit{73.8\%} \\
\midrule
A.I.G@GPT-5.4                 & 85.0\% & 79.7\% & 94.0\% & 86.3\% \\
A.I.G@Sonnet-4.6               & 88.5\% & 86.9\% & 90.6\% & 88.7\% \\
A.I.G@GPT-OSS-120B            & 69.7\% & 73.5\% & 61.5\% & 67.0\% \\
A.I.G@Qwen3-235B-A22B        & 86.3\% & 85.1\% & 88.0\% & 86.6\% \\
A.I.G@DeepSeek-R1-671B     & 85.9\% & 85.0\% & 87.2\% & 86.1\% \\
\textbf{A.I.G@Average}                       & \textit{83.1\%} & \textbf{82.0\%} & \textit{84.3\%} & \textit{82.9\%} \\
\midrule
SkillSpector@GPT-5.4                  & 69.2\% & 62.0\% & \textbf{99.1\%} & 76.3\% \\
SkillSpector@Sonnet-4.6               & 73.9\% & 66.5\% & 96.6\% & 78.7\% \\
SkillSpector@GPT-OSS-120B             & 57.7\% & 55.4\% & 78.6\% & 65.0\% \\
SkillSpector@Qwen3-235B-A22B          & 70.5\% & 67.1\% & 80.3\% & 73.2\% \\
SkillSpector@DeepSeek-R1-671B         & 70.5\% & 65.6\% & 86.3\% & 74.5\% \\
\textbf{SkillSpector@Average}         & \textit{68.4\%} & \textit{63.3\%} & \textit{88.2\%} & \textit{73.6\%} \\
\midrule
MalSkills@GPT-5.4                 & 58.5\% & 54.8\% & 97.4\% & 70.2\% \\
MalSkills@Sonnet-4.6              & 62.0\% & 56.9\% & 98.3\% & 72.1\% \\
MalSkills@GPT-OSS-120B            & 67.1\% & 61.1\% & 94.0\% & 74.1\% \\
MalSkills@Qwen3-235B-A22B         & 77.4\% & 70.8\% & 93.2\% & 80.4\% \\
MalSkills@DeepSeek-R1-671B        & 70.1\% & 65.4\% & 85.5\% & 74.1\% \\
\textbf{MalSkills@Average}        & \textit{67.0\%} & \textit{61.8\%} & \textit{93.7\%} & \textit{74.2\%} \\
\midrule
MASB@GPT-5.4                      & 57.7\% & 61.5\% & 41.0\% & 49.2\% \\
MASB@Sonnet-4.6                   & 59.4\% & 70.4\% & 32.5\% & 44.4\% \\
MASB@GPT-OSS-120B                 & 58.1\% & 65.6\% & 34.2\% & 44.9\% \\
MASB@Qwen3-235B-A22B              & 57.7\% & 64.1\% & 35.0\% & 45.3\% \\
MASB@DeepSeek-R1-671B             & 60.7\% & 69.8\% & 37.6\% & 48.9\% \\
\textbf{MASB@Average}             & \textit{58.7\%} & \textit{66.3\%} & \textit{36.1\%} & \textit{46.6\%} \\
\midrule
SkillFortify               & \textit{61.5\%} & \textit{57.2\%} & \textit{91.5\%} & \textit{70.4\%} \\
\midrule
SkillLeakBench               & \textit{65.4\%} & \textit{74.3\%} & \textit{47.0\%} & \textit{57.6\%} \\

\bottomrule
\end{tabular}
}
\end{table}


\begin{figure}[htbp]
\centering
\includegraphics[width=\linewidth]{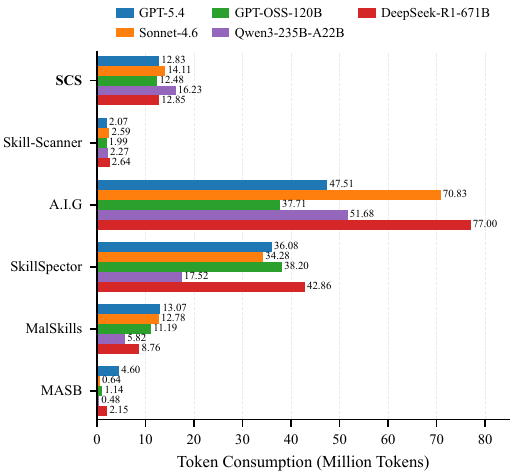}
\caption{Token Consumption Across Different Skill Detection Methods}
\label{fig:token_consumption}
\end{figure}

The detection results and LLM token consumption are summarized in Table~\ref{tab:cross_model_performance} and Figure~\ref{fig:token_consumption}, respectively.
Across the five evaluated LLMs, \textit{SCS} achieves an average accuracy of 85.6\%, precision of 80.2\%, recall of 95.2\%, and F1-score of 86.9\%, yielding the highest average accuracy, recall, and F1-score among the evaluated methods.
\textit{A.I.G} achieves a slightly higher average precision of 82.0\%, but its recall and F1-score decrease to 84.3\% and 82.9\%, respectively; moreover, its average token consumption is more than four times that of \textit{SCS} (56.9M vs. 13.7M).
\textit{SkillSpector} reaches a recall of 99.1\% with GPT-5.4, but this high recall does not translate into comparable overall performance because of its lower precision and F1-score, noticeable variation across LLMs, and substantially higher token consumption.
\textit{Skill-Scanner} and \textit{MalSkills} achieve average recalls of 89.7\% and 93.7\%, respectively, while consuming fewer tokens than \textit{SCS}; however, their substantially lower precision (62.8\% and 61.8\%) limits their average F1-scores to 73.8\% and 74.2\%.
\textit{MASB} also incurs relatively low token cost, but its low recall and F1-score indicate considerably weaker malicious-skill coverage.
Overall, \textit{SCS} achieves a favorable balance among detection effectiveness, robustness across different LLMs, and LLM token cost.

\textbf{Case Study and Baseline Analysis.}
We revisit the malicious \texttt{pptx\_skill} introduced in Section~\ref{sec:motivation} to examine how the seven baselines handle its cross-domain behavior. 
\textit{SkillFortify} and \textit{SkillLeakBench} misclassify this sample as benign. 
Both rely on predefined static rules, and their existing rules cannot reconstruct the dependency between the natural-language workflow and the invoked script, causing them to miss the complete attack path.
The LLM-based \textit{MASB}, \textit{MalSkills}, and \textit{Skill-Scanner} either fail to produce a malicious classification or exhibit inconsistent results across different LLMs. 
This model dependence highlights the uncertainty introduced when core semantic analysis and maliciousness classification rely heavily on LLM judgments. 
Although \textit{A.I.G} and \textit{SkillSpector} successfully detect the sample, their reports include some incorrect or irrelevant risk evidence. 
In contrast, \textit{SCS} consistently identifies the sample and reports focused evidence corresponding directly to the cross-domain attack path established in Section~\ref{sec:motivation}, further demonstrating the effectiveness of its unified cross-domain analysis.

\begin{tcolorbox}[top=0pt, bottom=0pt, left=3pt, right=3pt]
\textbf{Summary of RQ1:}
Across five LLMs, \textit{SCS} achieves the highest average accuracy
(85.6\%), recall (95.2\%), and F1-score (86.9\%) among the evaluated
methods while maintaining competitive precision (80.2\%).
Together with its moderate LLM token consumption, these results show
that \textit{SCS} provides a favorable balance between detection
effectiveness, cross-model robustness, and LLM usage.
\end{tcolorbox}

\subsection{RQ2: Quality of NL-to-IR Compilation}

The core stages of SCS form a structurally dependent analysis pipeline, where NL-to-IR provides the representation required for Text-PDG construction and subsequent cross-domain analysis. 
Removing an upstream stage would therefore make the downstream analysis undefined rather than yield a directly comparable reduced configuration. 
We instead perform stage-level validation of NL-to-IR, whose reliance on an LLM introduces potential errors that are not captured by system-level detection results alone. 
To this end, we construct a step-level gold set and compare the generated IR against human-annotated compilation requirements.

\subsubsection{Gold Set Construction}

Existing malicious-skill datasets are organized at the
skill-package level, whereas NL-to-IR compilation operates on
individual instruction steps. A skill-level label therefore
cannot determine whether a particular step has been correctly
compiled. We construct a dedicated step-level gold set to
evaluate this module.

The samples are drawn from a frozen snapshot of the ClawHub
corpus, which provides a large-scale collection of publicly
available real-world skills. We randomly sample 400 distinct
skills from this corpus and apply the deterministic
segmentation procedure described in
Section~\ref{sec:segmentation}. One non-empty
instruction step is then randomly selected from each skill, so
that no skill contributes more than one sample. 
This procedure avoids over-representing skills that contain more instruction steps.

Each sample contains a target instruction step, the preceding
context required to resolve its references, and a set of
expected compilation properties. We do not assign a single
fixed reference IR to each step, because the same instruction
may be compiled into different but semantically equivalent IR
statements, for example, by using different variable names or
decomposing one instruction into several operations. Instead,
we annotate a set of constraints that any correct compilation
should satisfy.

Specifically, \texttt{must\_call} specifies the API operations
that must appear, while \texttt{must\_not\_call} specifies
operations that must not be introduced without support from
the instruction. \texttt{must\_ground} requires concrete
arguments, such as file paths or URLs, to be supported by the
instruction or its context. \texttt{must\_thread} specifies the
producer--consumer dependencies that must be preserved between
operations, regardless of the concrete variable names used.
Finally, \texttt{must\_abstain} indicates that a step contains
no supply-chain-security-relevant operation and should
therefore generate no operational IR statement. A complete
annotation example is provided in
Appendix~\ref{app:nl_ir_gold_example}.

We construct a gold set containing 400 instruction steps.
This sample size is chosen because approximately 385
independent observations are required to estimate a binomial
proportion at a 95\% confidence level with a worst-case margin
of error of approximately $\pm5\%$. We round this number up
to 400 to provide sufficiently precise estimates of both the
overall step-level correctness and the individual property
satisfaction rates.

Two annotators independently label each sample based only on
the original instruction and its context, following a
predefined annotation guideline detailed in
Appendix~\ref{app:nl_ir_gold_guideline}. Each property is
assigned an explicit expected value, where the absence of a
required operation or dependency is also treated as a valid
expectation. The generated IR is then evaluated as satisfying
or violating each property, and a step is considered correct
only when all five properties are satisfied. We measure
inter-annotator agreement using Cohen's $\kappa$, resolve
disagreements through a third annotator, and report a 95\%
Wilson confidence interval for each result.

\subsubsection{Results}

The two annotators achieved Cohen's $\kappa$ values of 0.90, 0.93,
0.89, 0.84, and 0.89 for \texttt{must\_call},
\texttt{must\_not\_call}, \texttt{must\_ground},
\texttt{must\_thread}, and \texttt{must\_abstain}, respectively.
The macro-average Cohen's $\kappa$ was 0.89, indicating a high
level of agreement on the gold annotations.
Table~\ref{tab:nl_ir_accuracy} reports the compilation accuracy
for each annotated property across the five evaluated LLMs.
Across these models, all property accuracies remain above 94\%.
We further evaluate the overall step-level accuracy, where a step
is considered correct only when all five property requirements are
simultaneously satisfied. Under this stricter criterion, the evaluated
models achieve accuracies ranging from 92.5\% to 94.5\%, with a
cross-model difference of only 2.0 percentage points. 
The consistently high performance across different models indicates that the compilation pipeline is relatively insensitive to the choice of LLM.

\begin{table*}[t]
\centering
\caption{NL-to-IR compilation accuracy across LLMs on the gold set.
Each cell reports accuracy with a 95\% confidence interval.}
\label{tab:nl_ir_accuracy}
\small
\setlength{\tabcolsep}{4.5pt}
\renewcommand{\arraystretch}{1.4}
\resizebox{\textwidth}{!}{%
\begin{tabular}{@{}lcccccc@{}}
\toprule
\textbf{Model} &
\textbf{\texttt{must\_call}} &
\textbf{\texttt{must\_not\_call}} &
\textbf{\texttt{must\_ground}} &
\textbf{\texttt{must\_thread}} &
\textbf{\texttt{must\_abstain}} &
\textbf{Overall} \\
\midrule
GPT-5.4
& \shortstack{95.5\% (93.0--97.1)}
& \shortstack{99.2\% (97.8--99.7)}
& \shortstack{97.5\% (95.5--98.6)}
& \shortstack{99.5\% (98.2--99.9)}
& \shortstack{95.5\% (93.0--97.1)}
& \shortstack{92.5\% (89.5--94.7)} \\

Sonnet-4.6
& \shortstack{96.2\% (93.9--97.7)}
& \shortstack{99.8\% (98.6--100.0)}
& \shortstack{97.0\% (94.8--98.3)}
& \shortstack{100.0\% (99.0--100.0)}
& \shortstack{96.8\% (94.5--98.1)}
& \shortstack{93.8\% (90.9--95.7)} \\

GPT-OSS-120B
& \shortstack{95.0\% (92.4--96.7)}
& \shortstack{99.8\% (98.6--100.0)}
& \shortstack{98.0\% (96.1--99.0)}
& \shortstack{99.5\% (98.2--99.9)}
& \shortstack{97.0\% (94.8--98.3)}
& \shortstack{94.2\% (91.5--96.1)} \\

Qwen3-235B-A22B
& \shortstack{96.0\% (93.6--97.5)}
& \shortstack{99.8\% (98.6--100.0)}
& \shortstack{98.0\% (96.1--99.0)}
& \shortstack{100.0\% (99.0--100.0)}
& \shortstack{94.8\% (92.1--96.5)}
& \shortstack{94.5\% (91.8--96.3)} \\

DeepSeek-R1-671B
& \shortstack{95.5\% (93.0--97.1)}
& \shortstack{99.5\% (98.2--99.9)}
& \shortstack{97.8\% (95.8--98.8)}
& \shortstack{99.5\% (98.2--99.9)}
& \shortstack{95.8\% (93.3--97.3)}
& \shortstack{94.0\% (91.2--95.9)} \\

\bottomrule
\end{tabular}%
}
\end{table*}

\begin{tcolorbox}[top=0pt, bottom=0pt, left=3pt, right=3pt]
\textbf{Summary of RQ2:}
We evaluate the NL-to-IR compiler, the key LLM-dependent stage underlying subsequent cross-domain analysis, using a step-level gold set constructed from 400 randomly sampled real-world ClawHub skills and five semantic compilation properties. 
Across five evaluated LLMs, SCS achieves 92.5\%--94.5\% overall step-level correctness, while all individual property accuracies exceed 94\%. 
These results indicate reliable and consistent NL-to-IR compilation across different LLMs.
\end{tcolorbox}

\subsection{RQ3: Real-World Malicious Skill Analysis}

We conduct a large-scale analysis of the real-world ClawHub ecosystem using a snapshot collected on 20 April 2026, which contains 66,157 publicly available skills.
Applying SCS to the complete corpus produces 2,473 candidate skills, each accompanied by one or more static evidence paths. 
To further validate these candidates, we exercise the candidate skills through the auxiliary dynamic validation procedure described in Appendix~\ref{app:dynamic_validation}, which collects runtime evidence from tool calls, file accesses, process executions, and network events.
We then manually review the static and runtime evidence together and ultimately confirm 627 malicious skills.
Dynamic validation serves only as supporting evidence, while the final maliciousness decisions are made through manual review.
The complete analysis results and related artifacts are publicly available in the artifact described in Appendix~\ref{app:open_science}.

The confirmed malicious skills are highly concentrated among a small
number of publishers: \texttt{hightower6eu} contributes 354 skills
(56.46\%), \texttt{sakaen736jih} contributes 198 (31.58\%), and
\texttt{zaycv} contributes another 20 (3.19\%), with the first two
publishers together accounting for 88.04\% of all confirmed cases.
We further examine the subsequent status of these malicious skills using registry records retrieved through ClawHub's official public API~\cite{openclaw_clawhub_api}. 
Notably, all malicious skills published by the two dominant publishers had been removed from the public ClawHub registry by the time of our follow-up check on 15 July 2026. 
Overall, only 23 of the 627 confirmed malicious skills remain identifiable, including 19 labeled malicious and four labeled suspicious. 
This substantial reduction suggests considerable temporal change in the availability of malicious skills within the public ecosystem.

\begin{tcolorbox}[top=0pt, bottom=0pt, left=3pt, right=3pt]
\textbf{Summary of RQ3:}
Applying SCS to 66,157 real-world ClawHub skills identifies 2,473 candidate skills, of which 627 are confirmed as malicious through auxiliary dynamic validation and manual review. 
These malicious skills are highly concentrated among a small number of publishers and show substantial temporal change in subsequent registry records: most are absent from the later registry, including all confirmed malicious skills associated with the two dominant publishers. 
\end{tcolorbox}

\section{Discussion}
\label{sec:discussion}

\subsection{Coverage of Security-Relevant Behaviors}
The current API set and security rules cover recurring behaviors
identified in software supply-chain and agent-skill security studies.
Previously unseen behaviors or unsupported library calls may remain
outside this coverage and lead to false negatives. Such behaviors can
be incorporated by adding the corresponding API definitions, TSC
transfer rules, and detection rules without changing the underlying
xPDG representation.

\subsection{Static Analysis Scope and Auxiliary Dynamic Validation}
As a pre-install static analyzer, SCS infers TSC properties from statically observable information in skill artifacts, which may not fully capture runtime-resolved resources, dynamically generated values, or execution context. 
For example, a known credential path can be classified as sensitive, whereas sensitive data stored under an uninformative or dynamically generated path may not be recognized.
In the real-world study, we apply auxiliary dynamic validation to every candidate skill identified by SCS, providing runtime evidence to support subsequent manual confirmation.
This procedure is not part of the core detection pipeline: RQ1 evaluates the static SCS pipeline alone, while dynamic validation is used only to support the final review in RQ3.

\subsection{LLM-Related Uncertainty}
NL-to-IR compilation and context-dependent judgments for ambiguous
instructions still rely on LLMs and may introduce omissions,
hallucinated operations, or incorrect dependencies. Although the
constrained IR, legality checks, and dependency reconstruction reduce
such errors, they cannot completely eliminate the uncertainty of
interpreting open-ended natural-language instructions.
\section{Related Work}

\subsection{Supply Chain Security in Software and LLM Agents}

\textbf{Traditional software supply chains.}
Software supply chain security has been extensively studied in ecosystems such as npm, PyPI, and RubyGems. Representative works detect malicious packages through metadata signals, static analysis, and dynamic analysis~\cite{ohm2020backstabber,zahan2022weak,duan2020towards}. A common assumption in this line of work is that the attack payload is ultimately expressed in executable code, such as installation scripts, import-time logic, or hidden runtime behaviors, and can therefore be analyzed with conventional code-centric techniques.

\textbf{Agent supply chain security.}
LLM agent supply chains expose a broader attack surface because third-party artifacts can influence behavior through both code and natural-language instructions. This risk is closely related to prompt injection and jailbreak-style manipulation~\cite{greshake2023not,zou2023universal,liu2023prompt}, making model-facing instructions themselves security-relevant. Recent studies characterize this emerging supply chain in terms of architecture, threat taxonomy, and trust boundaries~\cite{xu2026agent,li2026towards,jiang2026sok}, while others demonstrate practical attacks through skill injection, document-borne payloads, supply-chain poisoning, and automated exploitation~\cite{schmotz2026skill,jia2026skillject,qu2026supply,duan2026skillattack}. Together, these works show that agent supply chains extend conventional code-centric threats to hybrid instruction-and-code artifacts.

\subsection{Security Studies on Agent Skills}
\textbf{Adversarial studies.}
Recent studies show that agent skills can be exploited through diverse mechanisms. Skill-inject and SkillJect examine skill-based prompt injection, while supply-chain poisoning and SkillAttack study document-borne payloads and exploitation of latent vulnerabilities~\cite{schmotz2026skill,jia2026skillject,qu2026supply,duan2026skillattack}. SkillTrojan and BadSkill further demonstrate skill-level and model-in-skill backdoors~\cite{feng2026skilltrojan,tie2026badskill}, while other work reports credential leakage risks from cross-modal interactions among skill artifacts~\cite{chen2026your}. Dedicated benchmarks and large-scale empirical studies further characterize malicious skill behaviors in public marketplaces~\cite{liu2026malicious,liu2026agent}. Collectively, these studies highlight a multifaceted attack surface spanning prompt injection, supply-chain poisoning, vulnerability exploitation, and stealthy backdoors.

\textbf{Detection and auditing methods.}
Recent defense research has proposed several approaches for auditing malicious agent skills. Open-source tools like Caterpillar~\cite{AliceCaterpillar2026} and Skill-Scanner~\cite{CiscoSkillScanner2026} use rule-based and ML methods to detect risky behaviors.
Several papers extend this line with more advanced techniques. 
AGENTARMOR constructs a security program dependence graph (sPDG) from runtime tool invocations to detect unsafe behaviors~\cite{wang2025agentarmor}. MalSkills combines symbolic rules with LLM-assisted reasoning to extract Sensitive Security Operations (SSOs) and assess maliciousness~\cite{wang2026elementary}. 
SkillProbe uses multi-turn semantic auditing to verify whether a skill's behavior matches its declared capabilities~\cite{guo2026skillprobe}. 
SkillSieve integrates lightweight filters and LLM-based checks over intent, permissions, stealth, and consistency~\cite{hou2026skillsieve}. 
SkillFortify formalizes skill adversaries and dependency resolution in a unified framework~\cite{bhardwaj2026formal}. 
Additionally, MaliciousAgentSkillsBench offers a large-scale empirical study of malicious skills, helping benchmark malicious behaviors~\cite{liu2026malicious}. 
SkillClone focuses on clone detection and propagation in the skill ecosystem~\cite{zhu2026skillclone}. 
Malicious Or Not improves classification by incorporating repository context to reduce false positives~\cite{holzbauer2026malicious}.
Together, these works show the growing interest in pre-install skill auditing, with designs differing in runtime monitoring, LLM reasoning, and capability modeling.

\subsection{Static Analysis and Information-Flow Techniques}

Traditional static analysis tools such as Semgrep, CodeQL, and Joern~\cite{Semgrep2026,CodeQL2026,yamaguchi2014modeling} provide practical support for code-centric security auditing through pattern matching, relational querying, and code property graphs. FlowDroid~\cite{arzt2014flowdroid} further demonstrates the value of precise interprocedural source-to-sink reasoning in security analysis. More broadly, taint analysis and information-flow control provide widely used abstractions for modeling trust propagation and sensitive data flow~\cite{schwartz2010all,sabelfeld2003language,denning1976lattice,myers1997decentralized}, while IFDS/IDE offers a standard graph-based framework for interprocedural propagation~\cite{reps1995precise,sagiv1996precise}. Systems such as TaintDroid and DroidSafe~\cite{enck2014taintdroid,gordon2015information} further illustrate how precise flow modeling can uncover leakage and misuse in complex applications. Although these techniques were not originally designed for hybrid agent artifacts, they provide important building blocks for our SCS, including graph construction, source-to-sink reasoning, and structured flow propagation.

\subsection{Natural-Language Procedural Representation}

Prior work has explored translating natural language into structured
program representations. NL2Bash maps natural-language requests to
executable commands~\cite{lin-etal-2018-nl2bash}, while PAGED and
related procedural-graph methods recover actions and their logical
dependencies from documents~\cite{du-etal-2024-paged,ying-etal-2026-multi}.
WorkflowLLM and the FLOW-GEN approach introduced in FLOW-BENCH represent natural-language workflows using Python-style
intermediate programs~\cite{fan2025workflowllm,duesterwald-etal-2025-flow}.
BREX and ExIde further study business-rule dependencies through structured and executable grounding~\cite{yang-etal-2026-business}.
Inspired by these designs, SCS adopts a constrained program-like IR,
but specializes it for static security analysis by explicitly modeling
security-relevant operations, arguments, and cross-step data dependencies.

\section{Conclusion}

In this paper, we present SCS, a static analysis framework for
auditing the supply-chain security of LLM agent skills. SCS brings
natural-language instructions and bundled code into a unified
program-analysis pipeline through constrained NL-to-IR compilation,
cross-domain dependence modeling with xPDG, and TSC-based security
property propagation. This design enables SCS to reason about
security-relevant behavior chains that span heterogeneous skill
artifacts.

Across five LLMs, SCS achieves the highest average accuracy, recall,
and F1-score among the evaluated methods while maintaining a favorable
effectiveness--cost trade-off. Its NL-to-IR compiler correctly translates
over 92\% of human-annotated instruction steps. 
On 66,157 self-collected ClawHub skills, SCS identifies 2,473 candidate skills, of which 627 are confirmed as malicious through auxiliary dynamic validation and manual auditing, illustrating the practical applicability of program analysis to hybrid natural-language and code artifacts.





\section*{Ethics Considerations}

Our study analyzes only publicly available skill artifacts. Dynamic validation is conducted in disposable, isolated environments using synthetic files and nonfunctional canary credentials, without exposing real user data or production credentials. Potentially malicious findings are manually reviewed before any disclosure.





\bibliographystyle{IEEEtran}
\bibliography{mybib}

%
\appendix

\subsection{Generative AI Usage}
We used GPT-5.4 for minor linguistic refinements, including grammar correction and improvements to expression clarity. 
We also used Claude Code to assist with implementing parts of the experimental codebase and related analysis scripts. 
All AI-assisted code, experimental outputs, and summarized results were manually reviewed by the authors. We validated the correctness of the implementation and results through code inspection, repeated experiment execution, and consistency checks on the generated data and reported conclusions.

\subsection{Open Science}
\label{app:open_science}

All artifacts associated with this paper are available at the following anonymous repository:
\begin{verbatim}
https://anonymous.4open.science/r/scs-F00A
\end{verbatim}

The repository includes:

\begin{itemize}
    \item \textbf{Implementation:}
    The implementation of \textit{SCS}, including the NL-to-IR compilation pipeline, xPDG construction, TSC property propagation, and security-rule evaluation.

    \item \textbf{Evaluation Artifacts:}
    The datasets, experimental configurations, and results used in the evaluation, including the benchmark comparison, NL-to-IR compilation evaluation, and real-world auditing analysis.
\end{itemize}






\subsection{Node and Edge Types in xPDG}
\label{app:xpdg_schema}

Table~\ref{tab:xpdg_node_edge_types} summarizes the node and edge
types used in xPDG.
The node types represent the hierarchical and executable units of a
skill, including skills, functions, statements, parameter and return
interfaces, and control predicates.
The edge types capture both intra-domain program dependencies and
cross-domain connections, including containment, invocation, data
flow, parameter binding, return binding, and control dependence.
Together, these elements allow xPDG to preserve the program structure
within each domain while explicitly connecting data and control
dependencies between natural-language instructions and bundled code.

\begin{table}[ht]
\centering
\small
\caption{Node and edge types in xPDG.}
\label{tab:xpdg_node_edge_types}
\begin{tabular}{l l l}
\toprule
\textbf{Kind} & \textbf{Type} & \textbf{Description} \\
\midrule
\multirow{5}{*}{Node}
& \texttt{SKILL}
& analysis root \\

& \texttt{FUNCTION}
& function unit (both domains) \\

& \texttt{STMT}
& statement unit (both domains) \\

& \texttt{ACCESS\_PATH\_PORT}
& parameter / return interface \\

& \texttt{CONDITION}
& control predicate node \\

\midrule
\multirow{6}{*}{Edge}
& \texttt{CONTAINS}
& hierarchical inclusion \\

& \texttt{CALLS}
& invocation relation \\

& \texttt{DATA\_FLOW}
& def-use dependency \\

& \texttt{PARAM\_BINDING}
& argument to input binding \\

& \texttt{RETURN\_BINDING}
& return to caller binding \\

& \texttt{CONTROL\_DEP}
& control dependency \\
\bottomrule
\end{tabular}
\end{table}
\subsection{TSC Propagation Algorithm}
\label{app:tsc_algorithm}

Algorithm~\ref{alg:tsc_propagation} presents the complete
fixed-point procedure used to propagate TSC security properties over
the xPDG. 
Section~\ref{sec:tsc_transfer} describes how the output state of an
individual statement is derived from its input variables, API
security properties, argument-to-output transfer rules, and
literal-dependent refinements. 
Such local transfer rules are
sufficient for linear dependency chains, where each state can be
computed once in dependency order.

Real-world programs, however, may contain branching structures,
loops, interprocedural calls, and other cyclic dependencies. In these
cases, a state update at one node may affect statements that have
already been processed, and states arriving through multiple
predecessors must be conservatively combined. SCS therefore applies a
Kildall-style worklist algorithm~\cite{kildall1973unified} to perform
global propagation over the xPDG.
The complete
worklist-based propagation procedure is presented in
Algorithm~\ref{alg:tsc_propagation}.

\begin{algorithm}[h]
\caption{TSC Security Property Propagation}
\label{alg:tsc_propagation}
\LinesNumbered
\KwIn{
    xPDG \(G=(N,E)\); API properties
    \(\mathcal{P}
    =\{(B_f,\Pi_f,R_f)\mid f\in\mathcal{A}\}\)
}
\KwOut{
    fixed-point variable states \(\sigma\)
}

\ForEach{variable \(v\) in \(G\)}{
    \(\sigma[v] \gets \bot\)\;
}

\(N_S \gets \mathrm{StmtNodes}(G)\)\;
\(\mathit{worklist}
\gets \mathrm{ReversePostorder}(G,N_S)\)\;

\While{\(\mathit{worklist} \neq \emptyset\)}{
    \(s \gets \mathit{worklist}.\mathrm{dequeue}()\)\;
    \(f \gets \mathrm{InvokedAPI}(s)\)\;

    \If{\(f = \varnothing\) \textbf{or}
        \(\neg\mathrm{HasOutput}(s)\)}{
        \textbf{continue}\;
    }

    \(I \gets \bot\)\;

    \eIf{\(\Pi_f \neq \emptyset\)}{
        \ForEach{\(\langle p,D\rangle \in \Pi_f\)}{
            \(a \gets \mathrm{BindArgument}(s,p)\)\;
            \If{\(a\) is a variable}{
                \(I \gets I \sqcup
                \operatorname{proj}_{D}(\sigma[a])\)\;
            }
        }
    }{
        \ForEach{\(v \in \mathrm{InputVars}(s)\)}{
            \(I \gets I \sqcup \sigma[v]\)\;
        }
    }

    \(\tau \gets B_f \sqcup I\)\;

    \ForEach{\(r \in R_f\)}{
        \If{\(r.\mathrm{condition}(s,\sigma)\)}{
            \(\tau \gets r.\mathrm{adjust}(\tau, s, \sigma)\)\;
        }
    }

    \(o \gets \mathrm{OutputVar}(s)\)\;
    \(\tau \gets \sigma[o] \sqcup \tau\)\;

    \If{\(\tau \neq \sigma[o]\)}{
        \(\sigma[o] \gets \tau\)\;

        \ForEach{\(s' \in
        \mathrm{Dependents}(G,o)\)}{
            \If{\(s' \notin \mathit{worklist}\)}{
                \(\mathit{worklist}.\mathrm{enqueue}(s')\)\;
            }
        }
    }
}

\Return \(\sigma\)\;
\end{algorithm}

The algorithm initializes every variable to the lattice bottom and
processes statements in reverse postorder. For each statement, it
applies the transfer rules defined in Section~\ref{sec:tsc_transfer}
to compute the state of its output variable. Whenever an output state
changes, all dependent statements connected through def--use,
parameter-binding, return-binding, or other xPDG dependency edges are
re-enqueued for further processing. This allows newly derived
security information to propagate through both intra-domain and
cross-domain dependencies.

The procedure continues until no variable state changes. Because the
TSC domain is finite and every transfer operation is monotonic, each
state can only move upward in the lattice, and the analysis is
guaranteed to terminate at a fixed point. The resulting states
therefore conservatively summarize all security information
propagated through the xPDG.

\subsection{Limited IR API Security Properties}
\label{app:API_Security_Properties}
This section summarizes the default TSC security properties assigned
to the limited IR APIs.
Each API is assigned base Trust and Sensitivity properties for its
return value, together with the intrinsic Capability introduced by
the API invocation.
These properties serve as the initial inputs to the propagation
procedure described in Section~\ref{app:tsc_algorithm}; they may be
further refined by argument-to-output transfer rules and
literal-dependent conditions.

Table~\ref{tab:restricted-api-tsc} presents the complete security
properties of the limited IR APIs.
SCS additionally defines security properties for the supported
Python, JavaScript, and Shell functions used in Code-PDG construction.
Due to space limitations, the complete language-specific function
lists and their corresponding TSC properties are provided in our
publicly available artifact described in
Appendix~\ref{app:open_science}.
\begin{table*}[h]
\centering
\caption{Default TSC Security Properties of the Limited IR API Set}
\label{tab:restricted-api-tsc}
\small
\renewcommand{\arraystretch}{1}

\begin{tabular}{lllll}
\toprule
\textbf{Category} &
\textbf{API Functions and Parameters} &
\textbf{Trust} &
\textbf{Sensitivity} &
\textbf{Capability} \\
\midrule

\multirow{5}{*}{\textit{File System}}
  & \texttt{read\_file(path)}
  & \texttt{PROVISIONED} & \texttt{PUBLIC} & \texttt{READ} \\

  & \texttt{glob(pattern)}
  & \texttt{PROVISIONED} & \texttt{PUBLIC} & \texttt{READ} \\

  & \texttt{search(pattern, path)}
  & \texttt{PROVISIONED} & \texttt{PUBLIC} & \texttt{READ} \\

  & \texttt{write\_file(path, content)}
  & -- & -- & \texttt{WRITE} \\

  & \texttt{edit\_file(path, content)}
  & -- & -- & \texttt{WRITE} \\

\midrule

\multirow{2}{*}{\textit{Execution}}
  & \texttt{bash(cmd, *args)}
  & \texttt{TRUSTED} & \texttt{PUBLIC} & \texttt{EXEC} \\

  & \texttt{run\_script(path, *args)}
  & \texttt{TRUSTED} & \texttt{PUBLIC} & \texttt{EXEC} \\

\midrule

\multirow{9}{*}{\textit{Network}}
  & \texttt{http\_get(url)}
  & \texttt{PROVISIONED} & \texttt{PUBLIC} & \texttt{NETWORK} \\

  & \texttt{http\_post(url, *payloads)}
  & \texttt{PROVISIONED} & \texttt{PUBLIC} & \texttt{NETWORK} \\

  & \texttt{send\_email(to, subject, body)}
  & -- & -- & \texttt{NETWORK} \\

  & \texttt{scp(dest, payload)}
  & -- & -- & \texttt{NETWORK} \\

  & \texttt{sftp(dest, payload)}
  & -- & -- & \texttt{NETWORK} \\

  & \texttt{rsync(dest, payload)}
  & -- & -- & \texttt{NETWORK} \\

  & \texttt{ssh(dest, payload)}
  & -- & -- & \texttt{NETWORK} \\

  & \texttt{nc(dest, payload)}
  & -- & -- & \texttt{NETWORK} \\

  & \texttt{ncat(dest, payload)}
  & -- & -- & \texttt{NETWORK} \\

\midrule

\multirow{2}{*}{\textit{Environment}}
  & \texttt{get\_env(name)}
  & \texttt{DERIVED} & \texttt{INTERNAL} & \texttt{ENV} \\

  & \texttt{set\_env(name, value)}
  & -- & -- & \texttt{ENV} \\

\midrule

\textit{Package Management}
  & \texttt{install\_package(manager, package)}
  & -- & -- & \texttt{INSTALL} \\

\midrule

\multirow{4}{*}{\textit{Data Transformation}}
  & \texttt{encode(data, encoding)}
  & \texttt{TRUSTED} & \texttt{PUBLIC} & -- \\

  & \texttt{decode(data, encoding)}
  & \texttt{TRUSTED} & \texttt{PUBLIC} & -- \\

  & \texttt{parse\_json(text)}
  & \texttt{TRUSTED} & \texttt{PUBLIC} & -- \\

  & \texttt{extract(data, pattern)}
  & \texttt{TRUSTED} & \texttt{PUBLIC} & -- \\

\midrule

\multirow{3}{*}{\textit{Cross-Skill}}
  & \texttt{invoke\_tool(name, **kwargs)}
  & \texttt{UNTRUSTED} & \texttt{INTERNAL} & \texttt{ELEVATE} \\

  & \texttt{import\_skill(name)}
  & \texttt{UNTRUSTED} & \texttt{INTERNAL} & \texttt{ELEVATE} \\

  & \texttt{tool\_arg()}
  & \texttt{UNTRUSTED} & \texttt{SECRET} & -- \\

\midrule

\textit{User Input}
  & \texttt{user\_input(prompt)}
  & \texttt{TRUSTED} & \texttt{INTERNAL} & -- \\

\midrule

\textit{Agent Control}
  & \texttt{attempt\_policy\_override(nl)}
  & -- & -- & \texttt{CONTROL\_OVERRIDE} \\

\bottomrule
\end{tabular}

\vspace{3pt}
\parbox{\textwidth}{\footnotesize
\textit{Note:} Trust and Sensitivity report the API base state before
input-state propagation and conditional state increases. For APIs without a
return value, these two entries are denoted by ``--''. In the Capability column,
``--'' indicates that the API introduces no intrinsic capability.
}
\end{table*}

\subsection{NL-to-IR Gold-Set Annotation}
\label{app:nl_ir_gold}

\subsubsection{Sample Construction and Dataset Characteristics}
\label{app:nl_ir_sample}

We construct a manually annotated gold set containing 400
natural-language instruction steps for evaluating the NL-to-IR
compilation capability of \textit{SCS}. 
We randomly select 400 distinct skills from the collected ClawHub
corpus and then randomly choose one non-empty instruction step from
each skill.
This procedure yields 400 instruction steps from 400 distinct
skills, ensuring that no skill contributes more than one sample.

Table~\ref{tab:nl_ir_sample_distribution} summarizes the distribution
of the resulting gold-set samples.

\begin{table}[t]
\centering
\small
\caption{Distribution of the NL-to-IR gold-set samples.}
\label{tab:nl_ir_sample_distribution}
\renewcommand{\arraystretch}{1.10}
\begin{tabular}{llr}
\toprule
\textbf{Dimension} & \textbf{Stratum} & \textbf{Steps} \\
\midrule
\multirow{3}{*}{Language}
  & English & 274 \\
  & Chinese or mixed Chinese-English & 115 \\
  & Other languages & 11 \\
\midrule
\multirow{8}{*}{Construct}
  & List item & 134 \\
  & Multi-operation span & 56 \\
  & Code block & 48 \\
  & Heading & 44 \\
  & Single-operation span & 44 \\
  & Non-operational prose & 38 \\
  & Inline code & 21 \\
  & Ordinary paragraph & 15 \\
\midrule
\multirow{5}{*}{Instruction length}
  & 25--50 characters & 107 \\
  & 51--100 characters & 150 \\
  & 101--180 characters & 98 \\
  & 181--280 characters & 37 \\
  & More than 280 characters & 8 \\
\bottomrule
\end{tabular}
\end{table}

The collected instructions span diverse linguistic forms and document
structures. As shown in Table~\ref{tab:nl_ir_sample_distribution}, the
gold set includes list items, code blocks, headings, inline code,
ordinary paragraphs, and both single- and multi-operation instruction
spans.

The instruction length ranges from 25 to 476 characters, with a median
of 78.5 characters and an average of 97.5 characters. Most samples are
relatively concise, supporting clear step-level annotation, while the
longer samples retain more complex operations and contextual information.

The sampled instructions also cover nine behavioral categories,
including execution, file access, network communication, package
management, environment operations, cross-skill invocation, input
handling, data transformation, and non-operational cases. Together,
these characteristics provide a diverse test set for evaluating whether
NL-to-IR compilation preserves operations, concrete arguments, and data
dependencies across different forms of natural-language instructions.

\subsubsection{Annotation Contract}
\label{app:nl_ir_contract}

Each instruction unit is annotated using five properties that describe the
semantic requirements of NL-to-IR compilation:

\begin{itemize}
\item \texttt{must\_call}: required IR API invocations. Each element
represents a required API group, allowing equivalent alternatives when
applicable;
\item \texttt{must\_not\_call}: prohibited API invocations unsupported by
the instruction;
\item \texttt{must\_ground}: concrete literals that must be preserved,
including URLs, paths, commands, or identifiers;
\item \texttt{must\_thread}: required producer-consumer dependencies between
operations;
\item \texttt{must\_abstain}: whether the instruction should not generate
executable IR actions.
\end{itemize}

The final annotation set covers all five properties. These labels constrain
the required semantics while allowing different but semantically equivalent
IR implementations.

\subsubsection{Annotation Example}
\label{app:nl_ir_gold_example}

Consider the following instruction:

\begin{quote}
\emph{First read the \texttt{Timezone:} field from \texttt{USER.md},
and then extract the corresponding value.}
\end{quote}

The instruction requires reading the specified file and extracting the
corresponding field from the retrieved content. Its gold annotation is:

\small
\begin{verbatim}
{
  "must_call": [
    ["read_file"],
    ["extract"]
  ],
  "must_not_call": [
    "bash",
    "http_get",
    "http_post",
    "write_file"
  ],
  "must_ground": [
    "USER.md",
    "Timezone:"
  ],
  "must_thread": [
    {
      "from": ["read_file"],
      "to": ["extract"]
    }
  ],
  "must_abstain": false
}
\end{verbatim}
\normalsize

This annotation requires the compiler to generate both operations, preserve
the referenced file name and field identifier, and maintain the dependency
between the file-reading and extraction steps. The annotation specifies the
required semantic constraints without prescribing a specific IR program or
variable naming strategy.

\subsubsection{Annotation Procedure and Criteria}
\label{app:nl_ir_gold_guideline}

Two annotators independently label every sampled instruction using the same
annotation guideline. They are shown only the instruction and its surrounding
context and remain blind to compiler outputs, model predictions, and each
other's annotations.

Before labeling the complete sample, the annotators first annotate a shared
calibration set and use disagreements to refine the guideline. After
independent annotation, Cohen's $\kappa$ is calculated for each property.
Remaining disagreements are resolved through a third annotator to produce the final gold labels.

Annotators record only behaviors explicitly supported by the instruction
itself. An API is required only when the corresponding operation is expressed
in the text. 
Literals are recorded only when they are supported by the instruction or its provided context,
and producer-consumer dependencies are labeled only when the instruction
states or clearly implies that one operation uses the result of another.

Descriptive prose, headings, and documentation-only references are
labeled as abstention cases when they do not specify executable behavior.
If a property admits multiple reasonable interpretations during
independent annotation, the annotators record the ambiguity rather than
forcing a speculative label.

\subsection{Dynamic Validation of Real-World Findings}
\label{app:dynamic_validation}

To validate the malicious behaviors identified during the
real-world study, we develop an auxiliary dynamic validation
procedure. This procedure is not part of the primary detection
pipeline of \textsc{SCS}; instead, it is used to collect runtime
evidence for statically identified candidate skills. Its purpose is
to determine whether a suspicious behavior can be reproduced in a
controlled environment, rather than to replace the static analysis
or independently classify all skills.
A candidate skill may contain one or more static findings. Accordingly, while RQ3 reports results at the skill level, dynamic validation operates at the finding level over individual evidence paths.

\subsubsection{Validation Workflow}

The dynamic validation procedure takes as input a skill package,
the corresponding static findings, and the source and sink locations
identified by \textsc{SCS}. Each skill is installed and exercised
inside a disposable sandbox that contains a synthetic user
environment, controlled tools, and nonfunctional canary credentials.
No real user data, host credentials, or cloud permissions are exposed
to the analyzed skill.

Unlike conventional package sandboxes that only execute explicit
entry points, the validation environment includes an instrumented
agent that loads and uses the skill through its intended activation
mechanism. This is necessary because security-relevant behavior may
be triggered by natural-language instructions in
\texttt{SKILL.md}, rather than by directly executing a bundled
script. When a skill also contains executable components, the
procedure may additionally invoke its installation scripts or
identified program entry points.

The overall workflow is as follows:

\begin{itemize}
    \item \textsc{SCS} provides the suspicious source, sink, and
    cross-domain evidence path identified during static analysis.
    \item A validation task is constructed to exercise the skill in
    a realistic usage context related to the static finding.
    \item The skill and the agent are executed inside an isolated,
    short-lived sandbox containing synthetic files and canary values.
    \item Runtime file, process, tool-call, and network events are
    collected and correlated with the original static finding.
    \item Deterministic validation rules classify the finding as
    \textit{confirmed}, \textit{refuted}, or
    \textit{inconclusive}.
\end{itemize}

\subsubsection{Static-Guided Skill Execution}

Dynamic execution is guided by the results of the static analysis.
For each candidate finding, the validation procedure uses the
declared skill functionality, relevant instructions in
\texttt{SKILL.md}, and the source-to-sink path reported by
\textsc{SCS} to construct one or more triggering tasks. The tasks are
phrased as ordinary user requests rather than explicit attack
instructions.

For example, if static analysis reports that a skill may read an SSH
private key and send its content to an external endpoint, the
validation environment may ask the agent to perform a legitimate
private-repository task according to the skill's documented
instructions. A synthetic SSH key containing a unique canary value is
placed at the expected location. If the skill reads this file and the
canary subsequently appears in an outbound request, the execution
provides direct evidence supporting the static finding.

The triggering task may be generated from predefined templates or
with LLM assistance. However, task generation is used only to improve
path coverage. Neither the triggering model nor the executing agent
is allowed to determine whether the skill is malicious.

\subsubsection{Isolation and Runtime Evidence}

Each validation run is performed in a disposable isolated
environment with bounded CPU, memory, process, network, and execution
time. The skill is provided with a synthetic home directory and
workspace containing nonfunctional credential files, environment
variables, and project artifacts. Each sensitive canary is unique to
one validation run, allowing any observed use or disclosure to be
attributed to a specific skill and execution.

The environment records evidence at multiple levels:

\begin{itemize}
    \item \textbf{Agent and tool events}, including skill activation,
    tool invocations, normalized arguments, and execution outcomes;
    \item \textbf{Language and process events}, including script
    execution, subprocess creation, file access, environment access,
    and dynamically loaded components;
    \item \textbf{Network events}, including DNS queries, connection
    destinations, HTTP metadata, and request content when available.
\end{itemize}

Events generated by the analyzed skill are distinguished from
sandbox initialization and agent-runtime activity. The collected
events are associated with the validation task and the corresponding
static finding, enabling the runtime behavior to be compared with the
source, sink, and dependency path reported by \textsc{SCS}.

\subsubsection{Validation Decisions}

The final validation result is derived from runtime events and
predefined rules, rather than from the natural-language judgment of
the agent or an LLM. We use the following three outcomes:

\begin{itemize}
    \item \textbf{Confirmed}: the suspicious operation is reached and
    runtime evidence supports the security consequence identified by
    the static finding. Examples include a canary secret appearing in
    an outbound request, execution of a downloaded payload, or an
    unauthorized modification of a protected artifact.
    \item \textbf{Refuted}: the relevant static location is
    successfully exercised under complete monitoring, but the
    predicted security consequence is not observed.
    \item \textbf{Inconclusive}: the suspicious location is not
    reached, the agent refuses or fails to activate the skill, the
    execution times out, monitoring is incomplete, or the observed
    evidence is insufficient to support either confirmation or
    refutation.
\end{itemize}

Failure to reproduce a behavior is not treated as evidence that the
skill is benign. In particular, a finding cannot be marked as
refuted unless the corresponding operation is demonstrably reached
under a successful and sufficiently monitored execution. This
coverage-aware design prevents agent refusal, conditional triggers,
and incomplete sandbox execution from being incorrectly interpreted
as negative evidence.

The dynamic result is finally linked back to the original
\textsc{SCS} finding and used together with manual inspection when
confirming malicious skills in the real-world study.




\end{document}